\documentclass[11pt,a4paper]{article}
\usepackage{fullpage,graphicx,amssymb,amsmath,xspace,amsfonts,tabularx,clrscode}

\newcommand{\WD}{\ensuremath{\mathrm{WL}}\xspace}
\newcommand{\WRD}{\ensuremath{\WD\ensuremath_1}\xspace}
\newcommand{\WED}{\ensuremath{\WD_2}\xspace}
\newcommand{\WXD}{\ensuremath{\WD_\infty}\xspace}
\newcommand{\WBA}{\ensuremath{\mathrm{WBA}}\xspace}
\newcommand{\TBA}{\ensuremath{\mathrm{TBA}}\xspace}
\newcommand{\ABA}{\ensuremath{\mathrm{ABA}}\xspace}
\newcommand{\WBP}{\ensuremath{\mathrm{WBP}}\xspace}
\newcommand{\ABP}{\ensuremath{\mathrm{ABP}}\xspace}
\newcommand{\WOA}{\ensuremath{\mathrm{WOA}}\xspace}
\newcommand{\AOA}{\ensuremath{\mathrm{AOA}}\xspace}
\newcommand{\WOP}{\ensuremath{\mathrm{WOP}}\xspace}

\newcommand{\AXD}{\ensuremath{\mathrm{AD_\infty}}\xspace}
\newcommand{\ARD}{\ensuremath{\mathrm{AD_1}}\xspace}

\newcommand{\diam}{\mathrm{diam}}
\newcommand{\peri}{\mathrm{peri}}
\newcommand{\bbox}{\mathrm{bbox}}
\newcommand{\canon}{{\cal P}}
\newcommand{\refinement}{\mathrm{r}}
\def\ARRWW{$AR^2W^2$\xspace}

\def\andfrac#1/#2{%
   \leavevmode\kern.1em
   \raise.5ex\hbox{\the\scriptfont0 #1}\kern-.1em
   /\kern-.15em\lower.25ex\hbox{\the\scriptfont0 #2}}
\newcommand{\half}{\andfrac 1/2}
\newcommand{\quart}{\andfrac 1/4}
\newcommand{\twoth}{\andfrac 2/3}
\newcommand{\fivee}{\andfrac 5/8}

\newtheorem{theorem}{Theorem}
\newtheorem{lemma}{Lemma}
\newenvironment{proof}{Proof:}{\qed}
\def\squareforqed{\hbox{\rlap{$\sqcap$}$\sqcup$}}
\def\qed{\ifmmode\squareforqed\else{\unskip\nobreak\hfil
\penalty50\hskip1em\null\nobreak\hfil\squareforqed
\parfillskip=0pt\finalhyphendemerits=0\endgraf}\fi}

\begin{document}

\title{Locality and Bounding-Box Quality of Two-Dimensional Space-Filling Curves}
\author{%
Herman~Haverkort\thanks{Dept.\ of Computer Science, Eindhoven University of Technology, the Netherlands, cs.herman@haverkort.net}
\and
Freek~van~Walderveen\thanks{Dept.\ of Computer Science, Eindhoven University of Technology, the Netherlands, freek@vanwal.nl}
}
\maketitle

\begin{abstract}
Space-filling curves can be used to organise points in the plane into bounding-box
hierarchies (such as R-trees). We develop measures of the \emph{bounding-box quality} of
space-filling curves that express how effective different space-filling curves are for this
purpose. We give general lower bounds on the bounding-box quality measures and on
locality according to Gotsman and Lindenbaum for a large class of space-filling curves.
We describe a generic algorithm to approximate these and similar quality measures for any
given curve. Using our algorithm we find good approximations of the locality and the bounding-box
quality of several known and new space-filling curves. Surprisingly, some curves with relatively
bad locality by Gotsman and Lindenbaum's measure, have good bounding-box
quality, while the curve with the best-known locality has relatively bad bounding-box quality.
\end{abstract}

\section{Introduction}
A space-filling curve is a continuous, surjective mapping from $\mathbb{R}$ to $\mathbb{R}^d$.
It was not always clear that such a mapping would exist for $d > 1$, but in the
late 19th century Peano showed that it is possible
for $d=2$ and $d=3$~\cite{peano}. Since then, quite a number of space-filling
curves have appeared in the literature. Sagan wrote an extensive treatise on space-filling
curves~\cite{sagan}, which discusses most curves included in our study.
During the early days space-filling curves were primarily seen as a mathematical curiosity.
Today however, space-filling curves are applied in areas as diverse as load balancing for grid
computing, colour space dimension reduction, small antenna design, I/O-efficient computations on
massive matrices, and the creation of spatial data indexes. In this paper, we focus on the
application of space-filling curves to the creation of query-efficient
spatial data indexes such as R-trees.

\paragraph{Bounding-box hierarchies}
We consider the following type of spatial data indexes for points in the plane.
The data points are organised in blocks of at most $B$ points, for some parameter $B$, such that
each point is stored in one block. With each block we associate a bounding box, which is the
smallest axis-aligned rectangle that contains all points stored in the block. The block
bounding boxes are then organised in an index structure. \emph{Intersection queries} are answered as
follows: to find all points intersecting a query window $Q$, we query the index structure for all
bounding boxes that intersect $Q$; then we retrieve the corresponding blocks, and check the
points stored in those blocks one by one.
To find the \emph{nearest neighbour} to a query point~$q$, one can use the index to search blocks
in order of increasing distance from~$q$. Thus one retrieves exactly the blocks whose
bounding boxes intersect the largest empty circle around~$q$.

An R-tree~\cite{manolopoulos} is an example of the type of structure described above: the blocks
constitute the leaves of the tree, and the higher levels of the tree act as an index
structure for the block bounding boxes. In practice the query response time is mainly determined
by the number of blocks that need to be retrieved: this is because the bounding box index structure
can often be cached in main memory, while the blocks (leaves) with data points
have to be stored on slow external memory (for example a hard disk needing 10 ms for each seek).

\begin{figure}[tb]
  \centering
  \leavevmode\llap{\raisebox{30mm}{(a)}\quad}\includegraphics[height=33mm]{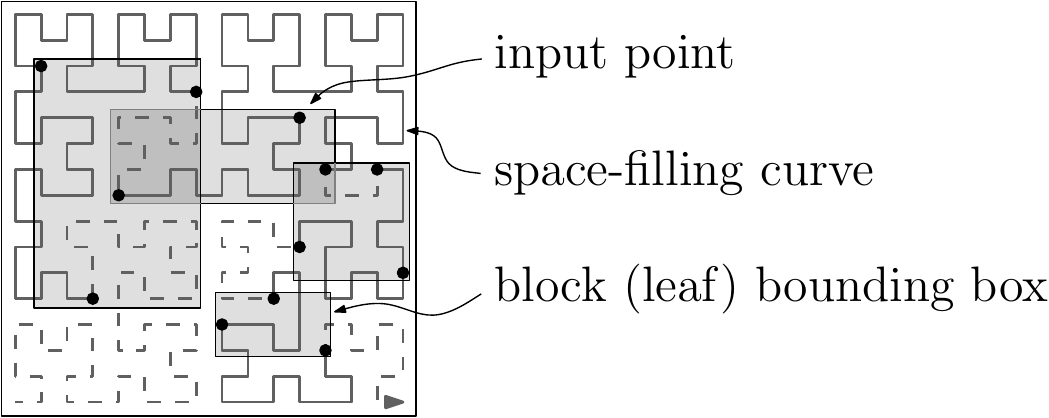}
  \quad
  \raisebox{30mm}{(b)}\quad\includegraphics[height=33mm]{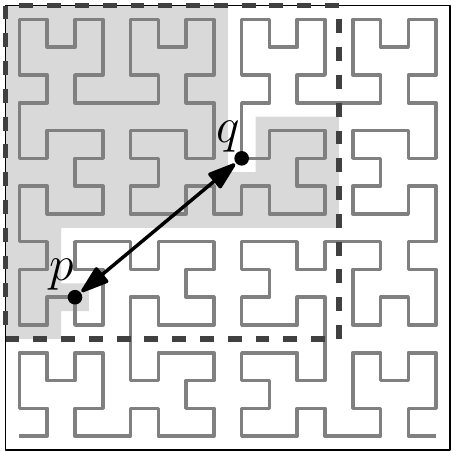}
  \caption{(a) Leaves of an R-tree with $B = 3$.\quad
  (b) Measuring locality for a particular curve section.
  $L_2$-locality ratio between $p$ and $q$ =
  squared Euclidean distance between $p$ and $q$, divided by the area covered by the curve section  between $p$ and $q$: $(6^2+5^2) / 87 \approx 0.70$.
  \quad
  Bounding-box area ratio between $p$ and $q$ =
  area of the bounding-box of the curve section $S$ between $p$ and $q$, divided by the area covered by $S$: $12\cdot 12 / 87 \approx 1.66$.
  (\WBA is the maximum over all pairs $p$ and~$q$)
  }
  \label{fig:rtree_leaves}
  \label{fig:measures}
\end{figure}

An R-tree is not uniquely defined by a set of data points. Any distribution of the input points
over the leaves (blocks) may be used as the basis of an R-tree, as long as each point is put
in exactly one block, and each block contains at most $B$ points. One way of making the distribution
is by ordering the input points along a space-filling curve~\cite{kamel1993} and then putting
each next group of $B$ points together in a block (see Figure~\ref{fig:rtree_leaves}(a)).

Since the number of blocks retrieved to answer a query is simply the number of bounding boxes
intersected, it is important that the ordering induced by the space-filling curve makes us fill
each block with points that lie close to each other and thus have a small bounding box. In fact
we can make some more precise observations for intersection queries with query windows that are
points or lines. For point-intersection queries we observe that if the data and query points lie
within a square of area~1, the average number of blocks retrieved for uniformly distributed point queries is simply the total area
of the bounding boxes.
For line-intersection queries with uniformly distributed orientation (between 0 and $\pi$) and signed distance
from the centre of the square (in an interval containing at least $[-\frac12\sqrt2, \frac12\sqrt2]$),
the chance of any particular block being retrieved is proportional to $\int_0^\pi w(\phi) d\phi$,
where $w(\phi)$ is the width of the bounding box in the orthogonal projection on a line with
orientation $\phi$. By the Crofton formula $\int_0^\pi w(\phi) d\phi$ is simply the perimeter of
the bounding box of the block, so for
uniformly distributed line queries, the average number of blocks
retrieved is proportional to the total perimeter of the bounding boxes. Therefore our goal is
to have bounding boxes with small (total) area and small (total) perimeter.

\paragraph{Our results}

We investigate which space-filling curves best achieve the above-mentioned goal:
sorting points into bounding boxes with small (total) area and small (total) perimeter.
To this end we propose new quality measures of space-filling curves that express how
effective different space-filling curves are in this context. We also provide an
algorithm to compute approximations of these and similar quality measures for any given curve.
We used this algorithm to compute approximations of known measures of so-called curve-to-plane
locality and of our new \emph{bounding-box quality} measures for several well-known and
new space-filling curves.

The known locality measures considered are the maximum, over all contiguous sections~$S$ of
a space-filling curve, of the squared $L_\infty$-, $L_2$- or $L_1$-distance between the endpoints
of $S$ divided by the area covered by $S$ (studied by Gotsman and Lindenbaum~\cite{gotsman} and
many other authors~\cite{alber,bauman,chochia,luxburg,niedermeier,niedermeier-manhattan}),
see Figure~\ref{fig:measures}(b) for an example.

Our first new measure is the maximum, over all contiguous sections~$S$ of a space-filling curve,
of the area of the bounding box of~$S$ divided by the area covered by~$S$. We call this measure
the \emph{worst-case bounding-box area ratio} (WBA, Figure~\ref{fig:measures}(b)).
Our second new measure considers 1/16th of the squared perimeter instead of the area, and we call
it \emph{worst-case bounding-box squared perimeter ratio} (WBP).

We prove that WBA and WBP are at least~2 for a large class of space-filling curves. We also
show that this class of curves has $L_2$-locality at least~4, thus complementing earlier
results by Niedermeier et al.~\cite{niedermeier} who proved this for another class of space-filling
curves (more restricted in one way, more general in another way).

We found that Peano's original curve achieves a WBA-value of less than 2.0001; the exact value
is probably exactly 2, which is optimal for this class of curves. Other well-known curves have
WBA-values ranging from 2.400 to 3.000. However, on the WBP measure Peano's curve is not that
good, with $\WBP = 2.722$.
Considering both WBA and WBP, the best curve we found in the literature is the
$\beta\Omega$-curve~\cite{wierum}, with $\WBA = 2.222$ and $\WBP = 2.250$. However, in this
paper we present a new variation on Peano's curve with even better scores:
a WBA-value of 2.000 and a WBP-value of 2.155. This variation also performs very well
on $L_\infty$-, $L_2$- and $L_1$-locality.

Both WBA and WBP consider the worst case over all possible subsections of the
curve. However, in the context of our application, it may be more relevant to study the
total bounding box area and perimeter of a set of disjoint subsections of the curve
that together cover the complete curve.
We can argue that in the limit, we may have the worst case for all subsections in such
a cover, but this seems to be unlikely to happen in practice.
Therefore we study the total bounding box area and perimeter of random
subdivisions of the curve into subsections. Here we find that many curves perform roughly
equally well, but those with particularly bad WBA- or WBP-values, such as the
Sierpi\'nski-Knopp curve~\cite{sagan} or H-order~\cite{niedermeier},
the \ARRWW-curve~\cite{asano},
or Peano's original (unbalanced) curve (regarding bounding box perimeters),
are clearly suboptimal in this sense.

We also estimate the total diameter of the subsections in random subdivisions of the curve
and present results for octagonal bounding boxes rather than rectangular bounding boxes.

Below we first explain how different space-filling curves can be described and how
they can be used to order points. We also describe some new curves.
Next we define the locality and bounding-box quality measures, and prove lower bounds.
After that we present our approximation algorithm, and present the results of our computations.

\section{Describing and using space-filling curves}

There are many ways to define space-filling curves, for example algebraic, like
in Peano's paper~\cite{peano}, or by describing an approximation of the curve by
a polyline, with a rule on how to refine each segment of the approximation
recursively~\cite{gardner}; many authors do this by specifying the regions filled
by sections of the curve together with the location of the endpoints of such
sections on the region boundaries (for example \cite{asano,sagan,wierum}).
Since we are concerned with the use of space-filling curves as a way to
order points in the plane, we choose a method to describe space-filling curves that is
based on defining how to order the space inside a (usually square) unit region.
We will see later how such a description is also a description of a curve.

\subsection{How to define and use a scanning order}
\label{howtouseit}

We define an order (\emph{scanning order}) $\prec$ of points in the plane as
follows. We give a set of rules, each of which specifies
(i) how to subdivide a region in the plane into subregions;
(ii) what is the order of those subregions; and
(iii) for each subregion, which rule is to be applied to establish the order
within that subregion.
We also specify a unit region of area 1 for each order (usually the unit square), and we
indicate what rule is used to subdivide and order it.
Technically it would be
possible to extend the orders to the full plane, but for simplicity we
rather assume that all data that should be ordered is first scaled to lie
within the unit region.

\begin{figure}[ht]
  \includegraphics[width=\textwidth]{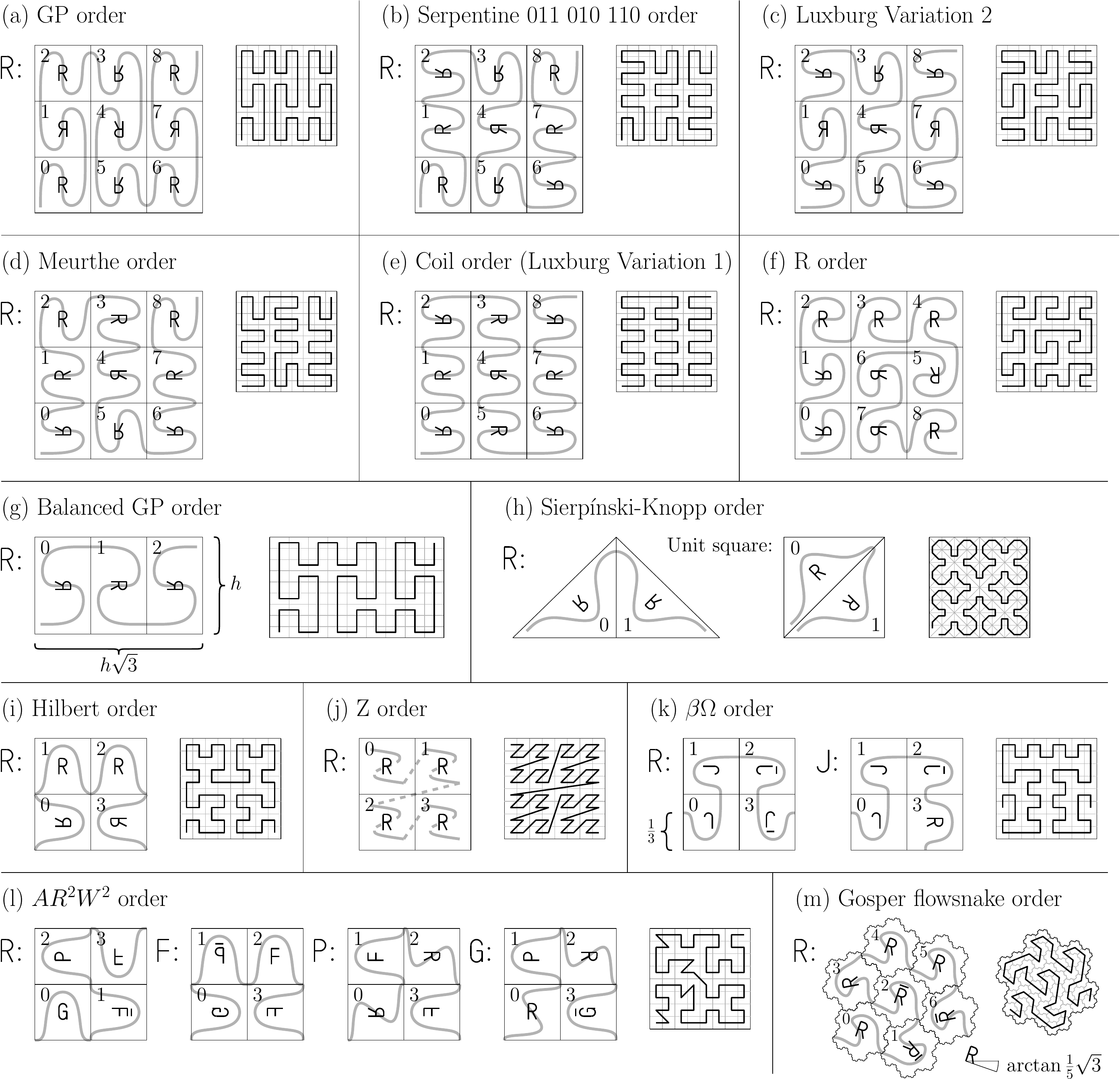}
  \caption{Space-filling curve definitions and example approximating polylines.}
  \label{fig:thecurves}
\end{figure}

The definitions of the scanning orders discussed in this paper are shown in Figure~\ref{fig:thecurves}.
Each rule is identified by a letter, and pictured by showing a region, its
subdivision into subregions, the scanning order of the subregions (by numbers \{0,1,2,...\}),
and the rules applied to the subregions (by letters). Variations of rules that
consist of simply rotating or mirroring the order of and within subregions,
are indicated by rotating or mirroring the letter identifying that rule.
Variations that consist of reversing the order of and within the subregion
are indicated by an overscore (Figure~\ref{fig:thecurves}(k,l,m))---making such
reversals explicit is the main difference between our notation and the notation of, for
example, Asano et al.~\cite{asano} or Wierum~\cite{wierum}.

We can now see how we can implement a comparison operator that allows us to sort
points according to a given scanning order. To decide whether $p$ precedes $q$
in the order, we determine in which subregions of the unit region $p$ and $q$
lie. If they are in different regions, $p$ precedes $q$ if and only if $p$ lies
in the lowest-numbered region of the two. If $p$ and $q$ lie in the same region,
we compare them according to the rule for that subregion recursively.
Ambiguity on the region boundaries can be resolved
as follows: horizontal and diagonal region boundaries are
always assumed to be included in the region above them; vertical boundaries
are assumed to be included in the region to the right.

Each drawing in Figure~\ref{fig:thecurves} includes a curve that
roughly indicates the scanning order within the subdivisions.
To obtain an
arbitrarily fine approximation of a space-filling curve corresponding to a
given scanning order, we may compute the subdivision of the unit region into
subregions recursively to the desired depth of recursion, and connect the centre
points of the resulting subregions by a polygonal curve in the order specified
by the rules.
Figure~\ref{fig:thecurves} includes a small example for each scanning order.

In the rest of this paper, whenever we write ``space-filling curve'', what we really
mean is the scanning order that defines it.

\subsection{The curves}
\label{thecurves}

The traditional scanning orders considered in this paper are the following.
\begin{itemize}
\item \textit{GP-order,} producing the space-filling
      curve described in detail by Giuseppe Peano~\cite{peano,moore} (Figure~\ref{fig:thecurves}(a)).
      We call this order \textit{GP-order} instead of \textit{Peano order} to avoid
      confusion with other curves that have also been referred to as the Peano curve by
      other authors.
\item \textit{Sierpi\'nski-Knopp-order,} producing the Sierpi\'nski-Knopp curve~\cite{sagan} (Figure~\ref{fig:thecurves}(h)).
      It orders triangular regions, and can be used to order points as
      described in Section~\ref{howtouseit}. Niedermeier et al.~\cite{niedermeier}
      describe how to use it to order squares in a $2^k \times 2^k$ grid for any
      $k \in \mathbb{N}$, their variation is called H-order.
      For all purposes in our paper, Sierpi\'nski-Knopp order and H-order are equivalent.
\item \textit{Hilbert order,} producing Hilbert's curve~\cite{hilbert} (Figure~\ref{fig:thecurves}(i)).
      One could say that Peano had already suggested that such a curve would exist~\cite{peano}.
\item \textit{Z-order,} which follows a space-filling curve by Lebesgue~\cite{lebesgue} (Figure~\ref{fig:thecurves}(j));
      Morton is credited with introducing Z-order to computer science~\cite{morton}.
\item \textit{Gosper flowsnake order}, producing the Gosper curve,
      also known as the flowsnake~\cite{akiyama,gardner} (Figure~\ref{fig:thecurves}(m)).
      It involves scaling with a factor $1/\sqrt{7}$ and rotations with angles of
      $\{0, \frac23\pi, \frac43\pi\} - \arctan{\frac15\sqrt3}$.
\end{itemize}
In addition to the above we include a number of variations on these orders.
Wunderlich~\cite{wunderlich} defined a class of orders that satisfy certain
restrictions, including:\begin{itemize}
\item \emph{simplicity}: the order is defined by only one rule, and transformed versions of it;
\item \emph{order-preservation}: transformations
      are rotations and/or reflections (no reversals);
\item \emph{edge-connectivity}: considering the set of regions obtained by applying the rule to any depth of
      recursion, we find that any two consecutive regions in the order share an edge;
\item \emph{uniformity}: all subregions have the same size.
\end{itemize}
Wunderlich categorises all different simple, order-preserving, edge-connected,
uniform orders based on subdividing the unit square in a grid of $2 \times 2$
or $3 \times 3$ squares (modulo rotations, reflections and reversals).
There is only one such order on a $2 \times 2$ grid (Hilbert order), and
there are only 273 on a $3 \times 3$ grid: one which we call \emph{R-order}\footnote{Wunderlich calls it \emph{Meander}, but
because other authors have used that name to describe other
orders, we rather use \emph{R-order}.} (Figure~\ref{fig:thecurves}(f)),
and 272 different orders that put the nine grid cells in the order of the `serpentine'
pattern of the GP-order (these differ in the rotations and reflections of the cells).
We investigated all of these orders to some extent, and based on the results we studied
some of them in more depth, particularly those depicted in Figures~\ref{fig:thecurves}(a)--(e).
By Wunderlich's numbering scheme these orders are \emph{Serpentine} orders
000\,000\,000, 011\,010\,110, 101\,010\,101, 110\,110\,110, and 111\,111\,111, respectively.
The first of these is simply GP-order.
Luxburg~\cite{luxburg} examined the third order calling it \emph{Variation~2};
we call the next order \emph{Meurthe order}
after the river in whose watershed we first presented it,
and we call the last order \emph{coil order}
(a variation also commonly found on the Internet, and studied by Luxburg as \emph{Variation~1}).
We explain in Section~\ref{results} why we chose exactly these orders.

When some of the above-mentioned restrictions are dropped, more orders are
possible. An example from the literature is Wierum's \emph{$\beta\Omega$-order}\footnote
{For simplicity we take an $\Omega$-shaped section of the curve. Wierum adds a special rule
for the unit square so that he gets a closed (cyclic) curve, but for the purposes of our
discussion this would be an unnecessary complication.}~\cite{wierum},
which is not simple, and unlike the other orders, it does not start and end in the
corners of the unit square (Figure~\ref{fig:thecurves}(k)). Another example is the \ARRWW-order which
is not simple and not edge-connected, but still \emph{vertex-connected} (Figure~\ref{fig:thecurves}(l)).
It was designed to have the special property that any axis-parallel square query window
can be covered by \emph{three} contiguous sections of the curve that together cover an area of at
most a constant times that of the query window~\cite{asano}---from most other curves in our
study one would need at least \emph{four} sections to get such a constant-ratio cover.

Most orders discussed so far cannot be scaled in only one dimension, because
their definitions involve rotations. For GP-order this is not the case.
In fact, as we will see later, a scaling in the horizontal dimension by a
factor $\sqrt{3}$ results in an order with much better locality properties.
We call the scaled order \emph{balanced GP-order} (Figure~\ref{fig:thecurves}(g)).

\section{Quality measures for space-filling curves}
\label{measures}

Several locality measures, or more generally, quality measures of space-filling curves
have been considered in the literature. These include:\begin{itemize}
\item bounds on the (average) distance between two points along the curve as a function
  of their distance in the plane~\cite{fishburn,mitchison,wierum-logarithmic-path-length}
  (non-trivial worst-case bounds are not possible~\cite{gotsman});
\item bounds on the (worst-case or average) distance between two points in the plane as
  a function of their distance along the curve~\cite{faloutsos,gotsman,niedermeier};
\item bounds on the (worst-case or average) number of contiguous sections of the curve that
  is needed to cover an axis-parallel query window, without covering too much space outside
  the query window~\cite{asano,faloutsos,jagadish,moon};
\item bounds on the (worst-case or average) perimeter or diameter of sections of the curve
  as a function of their area~\cite{hungershoefer,wierum}.
\end{itemize}
Not all methods of analysis considered in the literature can easily be applied to compare all curves
in our study. Some calculations or experiments in the literature are based on how the scanning
order sorts the points of a regular square grid whose size is a fixed integral power of the number of
subregions in the rule(s) that define(s) the order. Such measures may vary with the grid
size, which prevents a fair comparison between, for example, GP-order (for which we would
have to use a grid size that is a power of nine) and Hilbert order (for which we would have
to use a grid size that is a power of four). To enable a comparison between a broad range of
curves, we need measures that can be computed efficiently for large grids and converge when
the grid size goes to infinity.

\subsection{Notation}
Before we can discuss and analyse quality measures for space-filling curves in detail,
we need to introduce some notation.
For ease of writing, we assume for now that if a scanning
order is defined by more than one rule, then each rule contains the same number
of subregions.

A rule of a scanning order defines how to subdivide a unit region $C$ of size (area) 1
into $n$ subregions, numbered $0,...,n-1$. The scanning order inside subregion $i$ is
given by applying a transformation $\tau(i)$ to the unit region $C$ and the way it is
ordered by the ordering rule. For any base-$n$ number $a$ we use $a'$ to denote its first digit,
and $a''$ to denote the remaining digits. We use $C(a)$ as a shorthand for $\tau(a')(C(a''))$,
where $C(\emptyset) = C$. For example, $C(538)$ is subregion~8 of subregion~3 of
subregion~5, and it is found by applying transformation $\tau(5)$ to $C(38)$.
Similarly, we use $\tau(a)$ as a shorthand for $\tau(a') \circ \tau(a'')$,
where $\tau(\emptyset)$ is the identity transformation.

By $|A|$ we denote the size of a region $A$. We have $0 < |C(i)| < 1$ for all $0 \leq i < n$
(there are no empty subregions in the rules), and $\sum_{0 \leq i < n} |C(i)| = |C| = 1$.

Let $N_k$ denote the set of $k$-digit base-$n$ numbers.
We write $a \prec b$ if, in base-$n$ notation, $a$ and $b$ have the same number of digits
and $a < b$.
By $C(\preceq b)$ we denote the union of subregion $b$ and its predecessors, that is,
$\bigcup_{i=0}^{b'-1} C(i) \cup \tau(b')(C(\preceq b''))$, where $C(\preceq \emptyset) = C$.
Define $C(\prec b) := C(\preceq b) \setminus C(b)$,
$C(\succeq a) := C \setminus C(\prec a)$,
$C(\succ a) := C \setminus C(\preceq a)$, and
$C(a,b) := C(\prec b) \setminus C(\prec a)$.

Above we talked about the distance between two points along the curve, which may be a
somewhat counter-intuitive concept for a curve that can be refined and therefore lengthened
indefinitely. However, the distance between two points $p$ and $q$ along the curve is
well-defined as the area filled by the section of the curve that runs from $p$ to $q$,
or more precisely, as:
\[
|C(p,q)| := \lim_{k\to\infty}\quad \min_{a,b \in N_k \mbox{\ s.t.\ } p \in C(a), q \in C(b)} |C(a,b)|.
\]

\subsection{Pairwise locality measures}

From the quality measures mentioned above, the most relevant and applicable to the
construction of bounding-box hierarchies seem to be those that bound the (worst-case or
average) distance between two points in the plane as a function of their distance along the
curve. This is, intuitively, because points that lie close to each other along the curve are
likely to be put together in a block. Then, if the distance between those points in the
plane is small too, the block may have a small bounding box.

Gotsman and Lindenbaum~\cite{gotsman} defined the following class of locality measures:
\begin{equation*}
  \lim_{m\to\infty}\quad \max_{1 \leq i < j \leq m} \frac{d_r(S(i), S(j))^2}{(j - i)/m},
\end{equation*}
where $i$ and $j$ are integers, $S(i)$ is the $i$th square along the
curve in a subdivision of the unit square into a regular grid of $m$ squares, and $d_r(S, T)$ is
the $L_r$-distance between the centre point $(S_x,S_y)$ of $S$ and the centre point
$(T_x,T_y)$ of~$T$. Thus $d_r(S,T)=(|S_x-T_x|^r+|S_y-T_y|^r)^{1/r}$ for
$r \in \mathbb{N}$, and $d_\infty(S,T)=\max(|S_x-T_x|, |S_y-T_y|)$. We note that the measure is easily
generalised to scanning orders that are not based on regular grids, by defining it as:
\begin{equation*}
  \WD_r := \lim_{k\to\infty}\quad \sup_{a,b \in N_k} \frac{d_r(C(a),C(b))^2}{|C(a,b)|}.
\end{equation*}
We call this measure $\WD_r$ for \emph{Worst-case Locality}, as it indicates for points
that lie close to each other on the curve how far from each other they might get in the plane.
Since we have $d_1(p,q) \geq d_2(p,q) \geq d_\infty(p,q)$ for any pair of points
$p$ and $q$, we have $\WRD \geq \WED \geq \WXD$ for any space-filling curve.

\subsection{Pairwise bounding box measures}

Intuitively, one may expect a relation between locality and bounding box size, as
explained above. However, we may also try to measure bounding box size directly.
We define two measures to do so.
The first is the \emph{worst-case bounding box area ratio} (WBA):
\begin{equation*}
  \WBA := \lim_{k\to\infty}\quad \sup_{a,b \in N_k} \frac{|\bbox(C(a,b))|}{|C(a,b)|},
\end{equation*}
where $\bbox(S)$ is the smallest axis-aligned rectangle that contains $S$.
The second measure is the \emph{worst-case bounding box square perimeter ratio} (WBP):
\begin{equation*}
  \WBP := \frac1{16} \cdot \lim_{k\to\infty}\quad \sup_{a,b \in N_k} \frac{\peri(\bbox(C(a,b)))^2}{|C(a,b)|},
\end{equation*}
where $\peri(S)$ is the perimeter of $S$ in the $L_2$ metric. Taking the square of the perimeter
is necessary, because otherwise the measure would be unbounded as $k$ (the resolution of the ``grid'')
goes to infinity. Because the rectangle of smallest perimeter that has any given area is a square,
the ``ideal'' bounding box has squared perimeter 16. The division by 16 gives an easy relation
between \WBP and \WBA: we have $\WBP \geq \frac1{16} (4 \sqrt\WBA)^2 = \WBA$.
Furthermore, since the perimeter of the
bounding box of two points $p$ and $q$ is simply twice their $L_1$-distance, we have
$\WBP \geq \frac1{16} (2 \sqrt\WRD)^2 = \frac14 \WRD$.

We can define measures similar to \WBA and \WBP when the bounding boxes used are not axis-parallel rectangles,
but convex octagons whose sides have normals at angles of $0$, $\frac14\pi$, $\frac12\pi$, $\frac34\pi$,
$\pi$, $\frac54\pi$, $\frac32\pi$, and $\frac74\pi$ with the positive $x$-axis. We call these
measures \emph{worst-case bounding octagon area ratio} (\WOA) and \emph{worst-case bounding
octagon square perimeter ratio} (\WOP). In the definition of \WOP we still use the factor 16
to allow a direct comparison with \WBP and \WBA. Because the octagon of smallest perimeter
that has area 1 has squared perimeter $32/(1 + \sqrt 2) \approx 0.828 \cdot 16$, we have
$\WOP \geq 0.828 \cdot \WOA$.

\subsection{Total bounding box measures}

\paragraph{Worst-case}
For our application we argued that the average query response time is related to the total
area and perimeter of the bounding boxes formed by grouping data points according to a given
scanning order. When the points are sufficiently densely distributed in the unit region, the
gap in the scanning order between the last point of a group and the first point in the next
group will typically be small. Thus the grouping practically corresponds to subdividing the
complete unit region into curve sections, of which we store the bounding boxes.
To assess the quality of the order, we can define the \emph{worst-case total bounding box area} (TBA)
as follows:
\begin{equation*}
  \TBA := \lim_{k\to\infty}\quad  \sup_{a_1 \prec a_2 \prec ... \prec a_{m-1} \in N_k}
  \left(\sum_{i=1}^{m} |\bbox(C(a_{i-1},a_i))|\right),
\end{equation*}
where $a_0$ is defined as~0 and $a_m$ is defined as~$\emptyset$.
Since the bounding box area of each section $C(a_{i-1},a_i)$ is at most $\WBA \cdot |C(a_{i-1},a_i)|$
and the area of all sections together sum up to~1, the total bounding box area is clearly
at most~$\WBA$. But in fact we can prove the following.

\begin{lemma}
For any uniform scanning order, we have $\TBA = \WBA$.
\end{lemma}
\begin{proof}
\begin{figure}[tbp]
  \centering
  \includegraphics[width=\hsize]{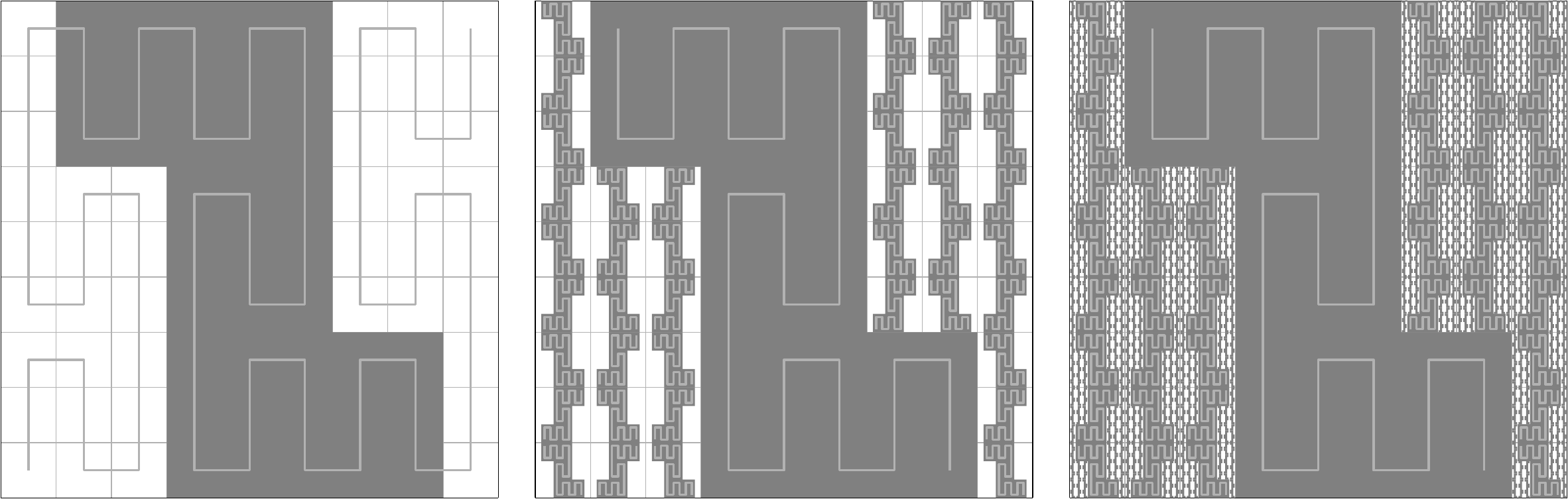}
  \caption{Left: an approximation of a section $S$ of the GP curve with $|\bbox(S)|/|S| = \WBA$. Middle and right: we can cover
           an arbitrarily large part of the unit region with such worst-case sections.}
  \label{fig:badtotal}
\end{figure}
Consider a section $S$ of the curve such that $|\bbox(S)|$ is equal to, or close to,
$\WBA \cdot |S|$. Now consider a recursive subdivision, following the rules of the
scanning order, of the unit region into a set $D$ of $m$ subregions (cells). Let $S'$ be
an approximation of $S$ that consists of all the cells of $D$ that are completely covered
by $S$, and let $D'$ be the remaining cells of $D$. Note that we can repeat the construction
within each of the cells of $D'$ (Figure~\ref{fig:badtotal}).
\begin{equation*}
\TBA \geq |\bbox(S')| + |D'|/m \cdot \TBA \geq |\bbox(S')| + (1-|S|) \cdot \TBA.
\end{equation*}
By choosing $m$ big enough we can let $|\bbox(S')|$ be arbitrarily close
to $\WBA \cdot |S|$. Thus we get:
\begin{equation*}
\TBA \geq \WBA \cdot |S| + (1 - |S|) \cdot \TBA,
\end{equation*}
which solves to $\TBA \geq \WBA$.
\end{proof}

\paragraph{Average}
As we showed above, the worst-case total bounding box area is not very informative.
Of greater practical relevance may be the \emph{average total bounding box area} (\ABA):
\begin{equation*}
  \ABA := \lim_{k\to\infty}  \mathrm{avg}_{a_1 \prec a_2 \prec ... \prec a_{m-1} \in N_k}
  \left(\sum_{i=1}^{m} |\bbox(C(a_{i-1},a_i))|\right),
\end{equation*}
where the average is taken over sets of $m-1$ cutting points $a_1, ..., a_{m-1}$ uniformly chosen from
the unit region. The above equation may serve as a complete definition of the \ABA measure for
\emph{fixed}~$m$, but this is not completely satisfactory. Experimental results with the scanning
orders described in this paper indicate that asymptotically, the measure does not grow or shrink
with $m$, but it exhibits a small fluctuation which repeats itself as $m$ increases by a
factor 3, 4 or 9 (depending on the curve)---in other words, the measure tends to be periodic
in $\log m$. Therefore any fixed choice of $m$ is likely to give an advantage to some scanning
orders and a disadvantage to others. Therefore, we define the average total bounding box area
more precisely as the above measure,
averaged over a range of values of $m$, such that $m$ is
large enough and $\log m$ is uniformly distributed in a range that covers an integral number
of periods of fluctuation of the curve under consideration.
We define an \emph{average total bounding octagon area} (\AOA) analogously.

We could define an \emph{average total bounding box square perimeter} in a similar way.
However, we are ultimately interested in the average perimeter, not the square perimeter.
We have to be more careful with the effect of $m$ now: we cannot expect to keep roughly
constant total bounding box perimeter as $m$ increases. To cut up a unit region into
$m$ sections such that their total bounding box perimeter is minimum, we would have to
cut it up into squares of area $1/m$ each, and their total
bounding box perimeter would be $4 \sqrt m$. Therefore the total bounding box perimeter
should be considered relative to $4 \sqrt m$, and we define the
\emph{square average relative total bounding box perimeter} (\ABP) as:
\begin{equation*}
  \ABP := \lim_{k\to\infty}  \left(\mathrm{avg}_{a_1 \prec a_2 \prec ... \prec a_{m-1} \in N_k} \frac{1}{4\sqrt m}
  \left(\sum_{i=1}^{m} \peri(\bbox(C(a_{i-1},a_i)))\right)\right)^2
\end{equation*}
In the above definition we still take the square in the end, to allow a direct comparison
between \ABP and \WBP.
The reader may verify that we must now have
$1 \leq \ABA \leq \WBA$ and $1 \leq \ABP \leq \WBP$.

\subsection{Total diameter measures}

Because for some applications it may be interesting to keep the diameter of curve
sections small~\cite{wierum} and because our software was easy to adapt to it, our
results in Section~\ref{results} also include estimations of the
\emph{square average relative total curve section diameter} (\AXD), defined as:
\begin{equation*}
  \AXD := \lim_{k\to\infty}  \left(\mathrm{avg}_{a_1 \prec a_2 \prec ... \prec a_{m-1} \in N_k} \frac{1}{\sqrt m}
  \left(\sum_{i=1}^{m} \diam_{\infty}(C(a_{i-1},a_i))\right)\right)^2,
\end{equation*}
where $\diam_\infty(S)$ is the diameter of $S$ in the $L_\infty$-metric.
We also compute \ARD: the same measure based on the $L_1$-metric.

\section{Lower bounds}

\subsection{Worst-case bounding box area}

\begin{theorem}
Any scanning order with a rule that contains a triangle has $\WBA \geq 2$.
\end{theorem}
\begin{proof}
The area of a bounding rectangle of any triangle $\triangle$
is at least twice the area of $\triangle$.
\end{proof}

\begin{theorem}
Any scanning order based on recursively subdividing an axis-aligned rectangle
into a regular grid of rectangles has $\WBA \geq 2$.
\label{lowerboundWBArectangles}
\end{theorem}
\begin{figure}[tb]
  \centering
  \includegraphics[width=0.44\hsize]{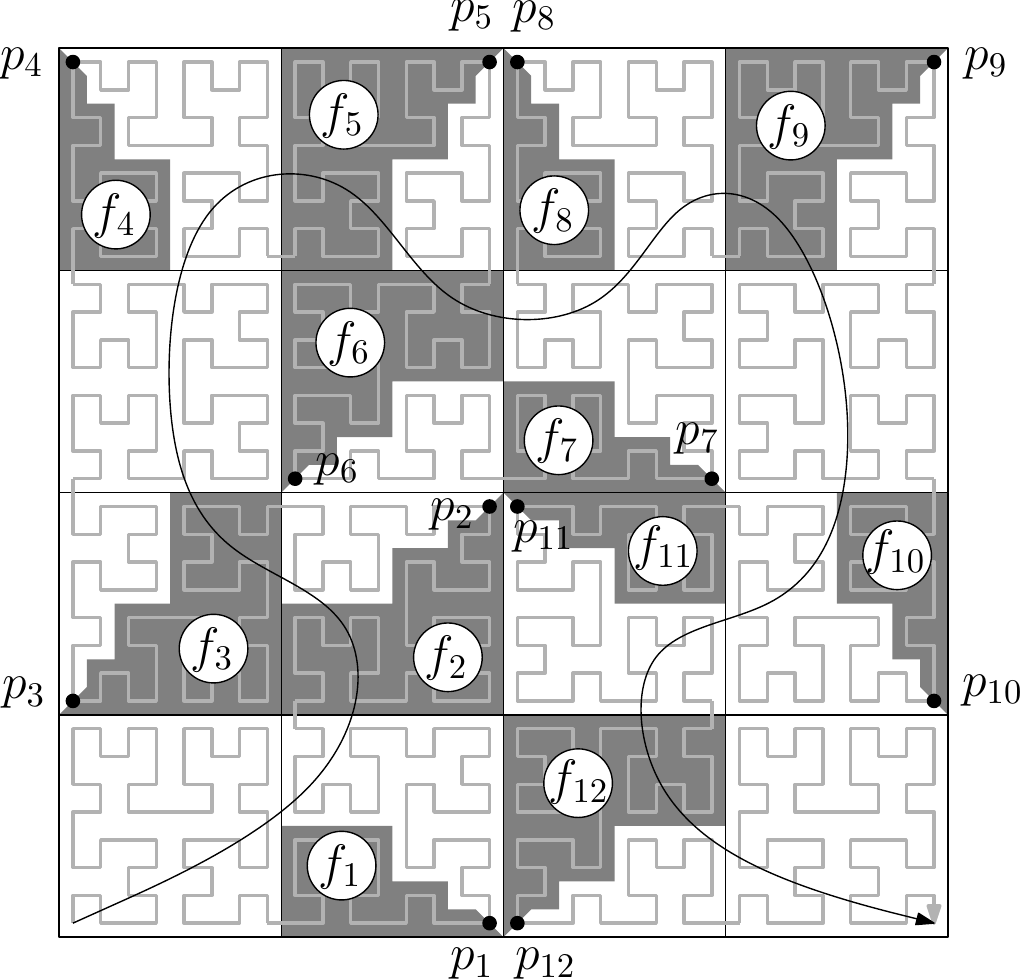}
  \caption{Corner rectangles in a grid. The smooth curve illustrates the order of the rectangles
           along the curve. In each corner rectangle, we marked the outer corner and we
           shaded the front part.}
  \label{fig:rectanglegrid}
\end{figure}
\begin{proof}
Consider a subdivision of the unit rectangle into a regular grid of
$m$~rectangles, following the rules of the scanning order recursively to the
depth where a grid of $\sqrt m \times \sqrt m$ rectangles is obtained.
We distinguish two cases: either there is a pair of rectangles that are
consecutive in the scanning order and do not share an edge, or all pairs of
consecutive rectangles share an edge.

In the first case, the bounding box of such a pair contains at least
four rectangles, and thus the curve section that covers that pair
results in $\WBA \geq 2$.

In the second case we argue as follows. Let
$s_1, ... s_m$ be these rectangles in order along the space-filling curve.
For each rectangle $s_i$ ($1 < i
\leq m$), let the edge of entry be the edge shared with $s_{i-1}$, and
for each rectangle $s_i$ ($1 \leq i < m$), let the edge of departure be
the edge shared with $s_{i+1}$. Among rectangles $s_2, s_3, ...,
s_{m-1}$, we distinguish two types of rectangles: straight rectangles and
corner rectangles.
A \emph{straight} rectangle is a rectangle whose edges of entry and
departure are not adjacent.
A \emph{corner} rectangle is a rectangle $s_i$ whose
edges of entry and departure share a vertex---we call this vertex the
inner corner of~$s_i$, and the opposite vertex is the outer corner of
$s_i$. The \emph{front part} of $s_i$ is the part of $s_i$ that appears
before the outer corner in the order.

Now we number the corner rectangles $t_1, t_2, ..., t_k$ in the order in which
they appear on the curve, let $p_1, p_2, ..., p_k$ be their outer
corners, and $f_1, f_2, ..., f_k$ be the areas of their front parts (Figure~\ref{fig:rectanglegrid}). Note
that any sequence of at least $\sqrt m$ rectangles must include a corner rectangle,
so $k \geq \sqrt m$. Consider the curve section from $p_i$ to $p_{i+2}$, for
any $i = 1, 2, ..., k-2$. Let the width of this section (by number of
rectangles) be $w$, let the height be $h$, and let $n \geq 3$ be the number of
rectangles from $t_i$ to $t_{i+2}$ inclusive. Observe that because
there is exactly one corner rectangle between $t_i$ and $t_{i+2}$, namely
$t_{i+1}$, we have $w \geq 2$, $h \geq 2$, and $w + h = n + 1$ (the $+1$
is because $t_{i+1}$ counts towards both $w$ and $h$). Now the area of
the curve section between $p_i$ and $p_{i+2}$ is $n-1 + f_{i+2} - f_i$,
and the area of its bounding box is $w \cdot h \geq 2(n - 1)$.
Hence we have $\WBA \geq 2(n-1) / (n - 1 + f_{i+2} - f_i)$, or equivalently,
$f_{i+2} - f_i \geq (2/\WBA - 1)(n-1) \geq 2\cdot (2/\WBA - 1)$.

For the sake of contradiction, suppose $\WBA < 2$. From the above we get
$f_{2i+2} - f_{2i} \geq 2\cdot (2/\WBA - 1)$ for all $i \in \{1,2,3,
..., m'\}$, where $m'$ is $\lfloor\frac12\sqrt{m}\rfloor - 1$. Therefore
$2/\WBA-1 \leq \frac1{2m'} \sum_{i=1}^{m'} (f_{2i+2} - f_{2i}) = \frac 1{2m'} (f_{2m'+2} - f_2) < \frac 1{2m'}$.
This must be true for any grid of rectangles that is obtained by refining the
subdivision recursively, following the rules of the scanning order.
So we must have $\lim_{m\to\infty} (2/\WBA - 1) = 0$ and thus $\lim_{m\to\infty} \WBA = 2$.%
\end{proof}

\subsection{Worst-case locality}

Niedermeier et al.~\cite{niedermeier} prove $\WED \geq \WXD \geq 3\half$ for scanning orders
that contain a section whose perimeter is an axis-aligned square.
The proof is based on
defining six points on the boundary of the square that need to be visited by the curve.
For each possible order in which these points may be visited, they add up the squares of the
distances between each pair of consecutive points in the order. Thus they derive a lower
bound on this sum that holds for all orders in which the points can be visited.
For the $L_\infty$- and $L_2$-metric this lower bound is at least $3\half$ times the area of
the square, and the lower bounds on $\WXD$ and $\WED$ follow.

Below we show how to apply this technique to triangular curve sections. Unfortunately it does
not work that well for rectangular curve sections. But there our new proof technique of
Theorem~\ref{lowerboundWBArectangles} comes to rescue, leading to better bounds.

\begin{theorem}
Any scanning order with a rule that contains a triangle has $\WED \geq 4$.
\end{theorem}
\begin{proof}
Consider a triangle that constitutes a contiguous section of the curve. Let $i_1$, $i_2$ and
$i_3$ be its three vertices, in the order in which they appear on the curve. Let $i_0$ be the
point on the edge $i_1i_3$ that appears before $i_2$ and is closest to $i_3$ among such points.
Let $i_4$ be a point on the segment $i_0i_3$, arbitrarily close to $i_0$. Let $w$ be the length
of the edge $i_1i_3$, and let $h$ be the height of the triangle relative to this edge, that is,
the distance between $i_2$ and the line containing $i_1i_3$, see Figure~\ref{fig:triangle}.

\begin{figure}
  \centering
  \includegraphics[scale=0.8]{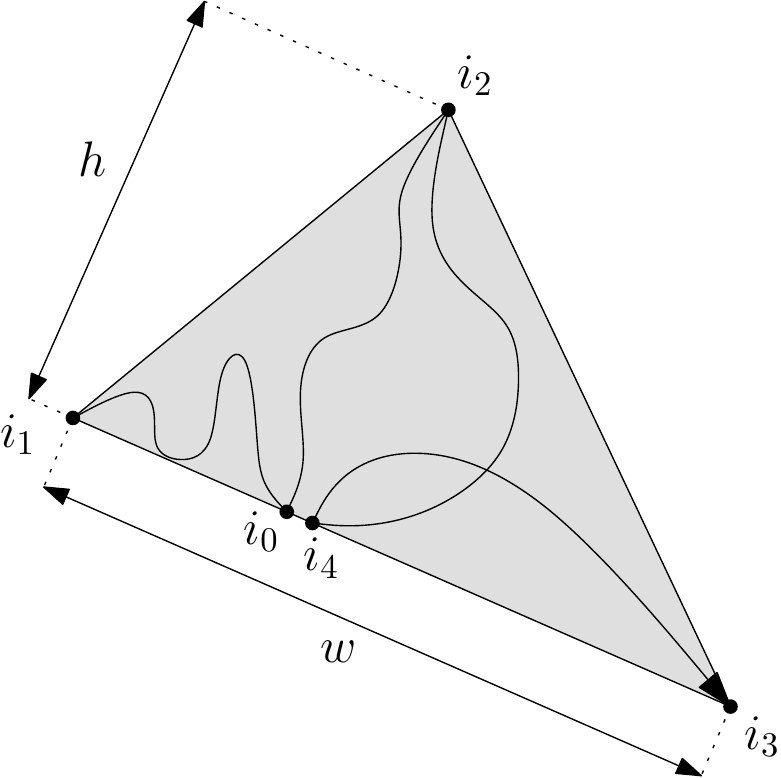}
  \caption{Definition of the points $i_0$, $i_1$, $i_2$, $i_3$ and $i_4$ on the boundary of a
           triangle, and a sketch of an order in which the space-filling curve may visit them.}
  \label{fig:triangle}
\end{figure}

The curve may pass through the points just defined in four different orders, it must be
$i_0 i_1 i_2 i_3 i_4$,
$i_1 i_0 i_2 i_3 i_4$,
$i_0 i_1 i_2 i_4 i_3$, or
$i_1 i_0 i_2 i_4 i_3$.

We will now analyse the last possibility in detail (the others can be checked in a similar
way). Consider the first leg of this path. When going from $i_1$ to $i_0$, the area
$C(i_1,i_0)$ filled by the curve must satisfy $d_2^2(i_1, i_0) / |C(i_1,i_0)| \leq \WED$,
that is:
\begin{eqnarray*}
|C(i_1,i_0)| & \geq & d_2^2(i_1, i_0) /\WED.
\end{eqnarray*}
Similarly we get:
\begin{eqnarray*}
|C(i_0,i_2)| \geq  d_2^2(i_0, i_2) /\WED & \geq &  h^2 /\WED,\\
|C(i_2,i_4)| \geq  d_2^2(i_2, i_4) /\WED & \geq &  h^2 /\WED,\\
|C(i_4,i_3)| \geq  d_2^2(i_4, i_3) /\WED & = &  \left(w^2 - 2w d_2(i_1, i_0) + d_2^2(i_1, i_0)\right) /\WED.
\end{eqnarray*}
Adding these up we get:
\[
|C(i_1,i_3)| \geq  \left(2 h^2 + w^2 - 2 w d_2(i_1, i_0) + 2d_2^2(i_1, i_0)\right) /\WED \geq  \left(2 h^2 + w^2 / 2\right)/\WED.
\]
Note that $C(i_1,i_0) \cup C(i_0,i_2) \cup C(i_2,i_4) \cup C(i_4,i_3) \cup C(i_1,i_3)$ is
at most the complete triangle, with $|C(i_1,i_3)| \leq hw/2$. Thus we get
$hw/2 \geq (2 h^2 + w^2 / 2) / \WED$, which solves to $\WED \geq 4h/w + w/h \geq 4$.

The other possible orders of $i_0, i_1, i_2, i_3$ and $i_4$ can be analysed in a similar way,
leading to the same result.
\end{proof}

Note that in fact we get $\WED \geq 4\alpha + 1/\alpha$, where $\alpha$ is the height/width
ratio that minimises the right-hand side. Somewhat surprisingly, this implies that optimal
locality cannot be achieved with equilateral triangles: with $\alpha = \frac12 \sqrt 3$ they
are subject to a lower bound of $\WED \geq 8/\sqrt 3$, while the triangles of the Sierpi\'nski-Knopp order,
with $\alpha = 1/2$, give $\WED = 4$~\cite{niedermeier}.

The proof technique of Niedermeier et al.\ could
also be modified for scanning orders that contain
rectangular (but not necessarily square) sections. However, in that case the lower bound will
drop below 3 (consider rectangles of aspect ratio $\sqrt 5$). Below we show how to use the
proof technique of Theorem~\ref{lowerboundWBArectangles} to get a lower bound of~4, not only
for \WED but also for \WXD.

\begin{theorem}
Any scanning order based on recursively subdividing an axis-aligned rectangle
into a regular grid of rectangles has $\WED \geq \WXD \geq 4$.
\end{theorem}
\begin{proof}
We follow the same approach as with Theorem~\ref{lowerboundWBArectangles}, but to get
a good bound on \WXD (not only \WED), we need to be a bit more careful in the definition
of the corners.

Consider a subdivision of the unit rectangle into a regular grid of
rectangles, following the rules of the scanning order recursively to the
depth where a grid of $\sqrt m \times \sqrt m$ rectangles is obtained.
Let $s_1, ... s_m$ be these rectangles in the order in which the
space-filling curve visits them. Note that each rectangle touches the next one,
at least in a vertex, otherwise \WED and \WXD would be unbounded.
Assume that the height/width ratio of each rectangle is $\alpha \geq 1$
(otherwise we swap the coordinate axes). Within this proof, we define the
width of a single rectangle to be $1/\sqrt\alpha$, and so its height is $\sqrt\alpha$
and its area is~1.

Among rectangles $s_2, s_3, ..., s_{m-1}$, we distinguish three types of rectangles.

A rectangle $s_i$ is \emph{straight} when $s_{i-1}$, $s_{i}$ and $s_{i+1}$ lie either in the
same row, or in three different rows.

A rectangle $s_i$ is a \emph{corner} rectangle when exactly one rectangle out of
$s_{i-1}$ and $s_{i+1}$ lies in the same row as~$s_i$. The \emph{outer corner} of~$s_i$ is
the vertex that lies farthest from $s_{i-1}$ and $s_{i+1}$; more precisely, if $s_{i-1}$
is in the same row as~$s_i$, the outer corner is the vertex of~$s_i$ that touches neither
the column of~$s_{i-1}$ nor the row of~$s_{i+1}$; otherwise, that is, if $s_{i+1}$ is in
the same row as~$s_i$, the outer corner is the vertex of~$s_i$ that touches neither
the row of~$s_{i-1}$ nor the column of~$s_{i+1}$.

A rectangle $s_i$ is a \emph{double corner} rectangle when $s_{i-1}$ and $s_{i+1}$ lie in
the same row, but not in the same row as $s_i$---implying that the curve makes something of
a U-turn in $s_i$. Such a rectangle $s_i$ has two outer corners, namely the vertices of
$s_i$ that do not touch $s_{i-1}$ or $s_{i+1}$. We distinguish a first corner and a second
corner, by the order in which they appear in the scanning order between $s_{i-1}$ and $s_{i+1}$
(Figure~\ref{fig:doublecorner}).

\begin{figure}
  \centering
  \includegraphics[scale=0.8]{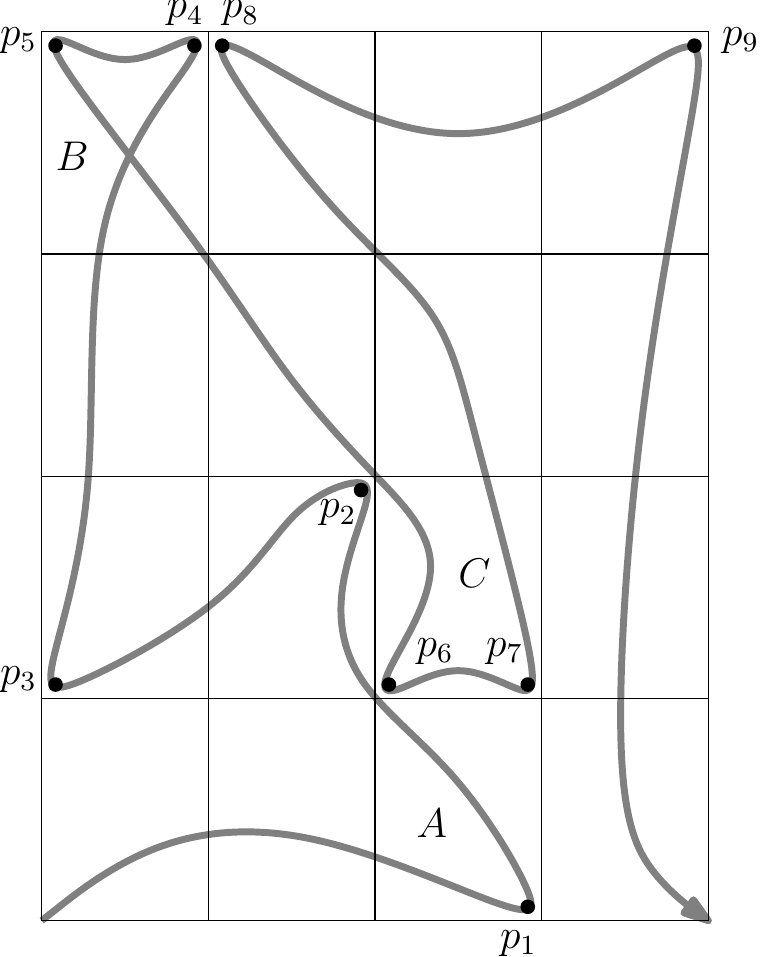}
  \caption{Corners of a scanning order that is not edge-connected on a rectangular grid.
           Note how horizontal and mainly vertical sections alternate between the corners.
           Although both the predecessor and the successor of rectangle $A$ lie to the left,
           it is \emph{not} a double corner rectangle, since the path to the predecessor is
           horizontal and a path to a rectangle in another row (such as the successor $A$)
           is always considered to be vertical. Rectangles $B$ and $C$ are double corner rectangles.}
  \label{fig:doublecorner}
\end{figure}

Now number the corner rectangles $t_1, t_2, ..., t_k$ in the order in which
they appear on the curve, with each double corner rectangle listed twice.
Let $p_1, p_2, ..., p_k$ be the outer corners of these rectangles. Where $t_i$ and $t_{i+1}$
are the two copies of a double corner rectangle, $p_i$ is the first corner and $p_{i+1}$ is
the second corner.
The \emph{front part} of $t_j$ is the part of $t_j$ that appears
before $p_j$ in the order.
As before, we have $k \geq \sqrt m$.

As before, we will argue about curve sections between two corners $p_i$ to
$p_{i+2}$, for even $i$. In addition we assume that for even $i$, the rectangles $t_i$ and
$t_{i+1}$ lie in the same row (if this is not the case, we could simply do the calculations
given below for odd $i$ instead of even $i$).

For the sake of contradiction, assume that $\WXD < 4$.

Let $w \geq 1$ be the number of rectangles in the order from $t_i$ to $t_{i+1}$ inclusive
($w$ could be~1 if $t_i = t_{i+1}$ is a double corner rectangle),
and $h \geq 2$ the number of rectangles in the order from $t_{i+1}$ to $t_{i+2}$ inclusive).

The $L_\infty$-distance between $p_i$ and $p_{i+1}$ is at least $w/\sqrt\alpha$,
hence we get\\
$\WXD \geq \frac 1\alpha w^2 / (w - 1 + f_{i+1} - f_i)$, which we rewrite as:
\begin{equation}\label{barconstraint}
f_{i+1} - f_i \geq \frac{w^2/\alpha}{\WXD} - (w - 1).
\end{equation}

The $L_\infty$-distance between $p_{i+1}$ and $p_{i+2}$ is at least $\sqrt\alpha h$.
Hence we get:
\begin{equation}\label{rawlegconstraint}
\WXD \geq \frac{\alpha h^2}{h - 1 + f_{i+2} - f_{i+1}},
\end{equation}
which we rewrite as:
\begin{equation}\label{legconstraint}
f_{i+2} - f_{i+1} \geq \frac{\alpha h^2}{\WXD} - (h - 1).
\end{equation}

Adding up Equations \ref{barconstraint} and~\ref{legconstraint} we get:
\begin{equation}\label{Lconstraint}
f_{i+2} - f_i \geq \frac{w^2/\alpha + \alpha h^2}{\WXD} - (w + h - 2).
\end{equation}
Furthermore Equation \ref{rawlegconstraint} gives us
$4 > \WXD \geq \alpha h^2 / (h - 1 + f_{i+2} - f_{i+1}) \geq \alpha h \geq h$.
From this we get that $h$ must be either 2 or 3.
In the case of $h = 2$ Equation \ref{Lconstraint} becomes:
\[
f_{i+2} - f_i \geq \frac{w^2/\alpha + 4 \alpha}{\WXD} - w = \left(\frac{w/\alpha + 4 \alpha/w}{\WXD} - 1\right) w \geq \frac{4}{\WXD} - 1.
\]
In the case of $h = 3$ and $w = 1$ Equation \ref{Lconstraint} becomes:
\[
f_{i+2} - f_i \geq \frac{1/\alpha + 9 \alpha}{\WXD} - 2 \geq \frac{10}{\WXD} - 2 > 2\left(\frac{4}{\WXD} - 1\right) > \frac{4}{\WXD} - 1.
\]
In the case of $h = 3$ and $w \geq 2$ Equation \ref{Lconstraint} becomes:
\[
f_{i+2} - f_i \geq
\frac{\frac{w^2}{\alpha} + 9 \alpha}{\WXD} - (w + 1) \geq
\frac{\frac{4(w+1)^2}{9\alpha} + 9\alpha}{\WXD} - (w + 1) =
\left(\frac{4\frac{w+1}{9\alpha} + \frac{9\alpha}{w+1}}{\WXD} - 1\right) (w + 1) \geq
\frac{4}{\WXD} - 1.
\]

Thus we get $f_{i+2} - f_i \geq 4/\WXD - 1$ in all cases.
The proof now concludes as for Theorem~\ref{lowerboundWBArectangles},
leading to the conclusion $\lim_{m\to\infty} \WXD = 4$, which also implies $\WED \geq 4$.
\end{proof}

Niedermeier et al.~\cite{niedermeier} also proved $\WED \geq \WXD \geq 4$, but for another class
of scanning orders, namely those that (i) contain a section whose perimeter is an axis-aligned square
and (ii) are cyclic, that is, the end of that section touches its beginning.
Our proof does not
need those conditions, but needs others. To prove $\WED \geq 4$ we require that the curve has a
triangular section, or that it has a rectangular section subdivided recursively into a regular
grid of rectangles.

Regarding $L_1$-locality, Niedermeier et al.\ proved $\WRD \geq 8$ if both
conditions (i) and (ii) hold, and
$\WRD \geq 6\half$ if only condition (i) holds. We have no results to complement this: it seems
hard to use our technique to prove any lower bound on $\WRD$ which is significantly better than 4.

\section{Approximating the worst-case measures}

In this section we describe how we can obtain upper and lower bounds on the
quality measures such as the worst-case locality and the worst-case bounding box quality
measures defined in Section~\ref{measures}.
For ease of description, we assume that the scanning order is defined by a
single recursive rule without reversals.
The techniques described below can easily
be extended to multiple-rule scanning orders or scanning orders with reversals
(in fact that is what we implemented).

Let $\mu$ be a mapping from regions to real numbers in a way that is
invariant under all transformations $\tau(i)$ involved in the
recursive definition of the scanning order. For example, $\mu(R)$ could be
$|\bbox(R)|/|R|$, or the square of the diameter of $R$ divided by~$|R|$.
Our goal is to approximate
$\mu^* = \lim_{k \rightarrow \infty} \sup_{i \prec j \in N_k} \mu(C(i,j))$.
(We may also let $\mu$ depend on $C(i)$ and/or $C(j)$.) The mapping $\mu$
must be well-defined when $C(i,j)$ is not empty; when $|C(i,j)| = 0$
we may assume $\mu(C(i,j)) = \infty$.

\subsection{Representing curve sections by probes}

We will compute the approximation of $\mu^*$ by exploring \emph{probes}. A probe $P$
is specified by three consecutive subsections of the order: a front section,
a midsection, and a tail section. The probe $P$ thus describes a set of
contiguous subsections of the scanning order, namely all those subsections $S$ that
start somewhere in the front section of $P$ and end somewhere in the tail section of $P$.
For any probe $P$, let $\alpha(P)$ be
the transformation that transforms $C$ into the front section of~$P$; let $M(P)$
be the midsection of~$P$; and let $\omega(P)$ be the transformation that transforms
$C$ into the tail section of~$P$. A section $P(i,j)$ of a probe $P$ is the region
$\alpha(P) C(\succeq i) \cup M(P) \cup \omega(P) C(\prec j)$.
Let $\mu^*(P)$ be the maximum value of $\mu(S)$ over all subsections $S$ covered
by the probe, that is, $\mu^*(P) = \lim_{k \rightarrow \infty} \sup_{i, j \in N_k} \mu(P(i,j))$.
A probe $P$ may be rotated, mirrored, scaled and/or reversed: this
does not affect the value of~$\mu^*(P)$.

\begin{figure}[tb]
  \centering
  \includegraphics[width=\hsize]{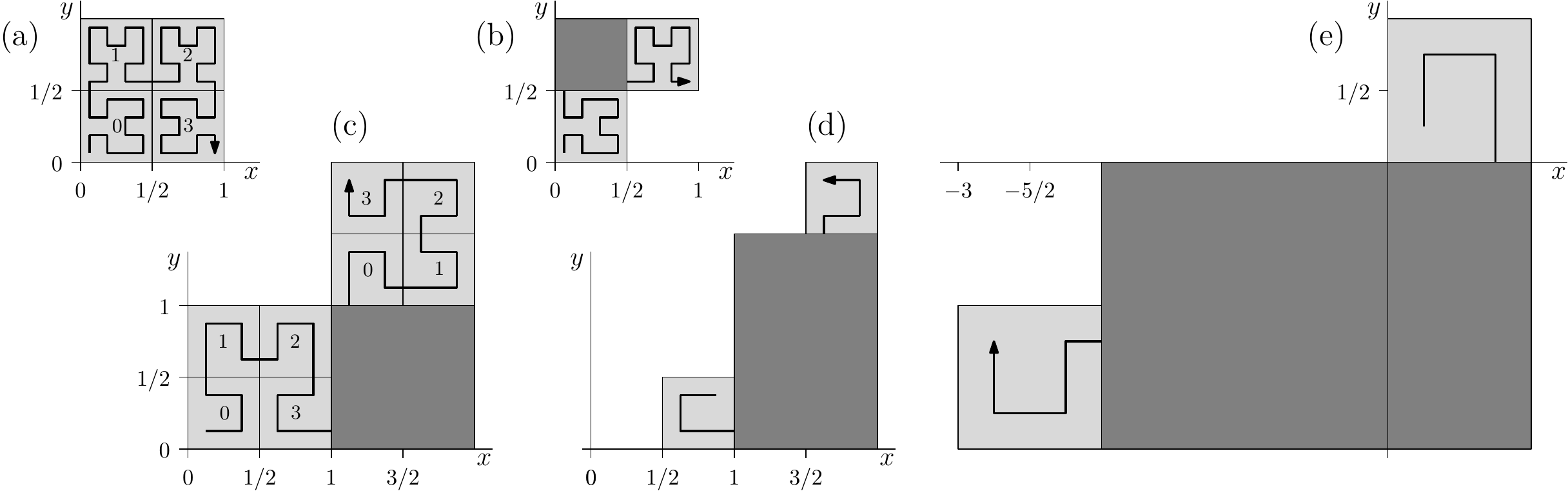}
  \caption{(a)~The Hilbert order.\quad
           (b)~Base probe~$B_{02}$ of the Hilbert order.
               The front transformation $\alpha(B_{02})$ consists of scaling with factor $1/2$ and a reflection in the line $x = y$.
               The midsection $M(B_{02})$ is the square $[0,\frac12] \times [\frac12,1]$ (the dark area in the figure).
               The tail transformation $\omega(B_{02})$ consists of scaling with factor $1/2$ and a translation over $(\frac12,\frac12)$.\quad
           (c)~Canonical form~$\canon$ of~$B_{02}$.
               The canonical form is obtained from $B_{02}$ by applying the transformation $\alpha(B_{02})^{-1}$, that is,
               reflection in the line $x = y$ and scaling with factor~2.\quad
           (d)~Refinement $\refinement_{32}(\canon)$ of~$\canon$.\quad
           (e)~The same refinement in canonical form: child~$\canon_{32}$ of~$\canon$.}
  \label{fig:probes}
\end{figure}
All subsections of the scanning order can be captured by a set of probes
as follows. For $0 \leq i < k < n$, let \emph{base probe} $B_{ik}$ be the
probe with front transformation $\tau(i)$, midsection
$\bigcup_{i < j < k} C(j)$, and tail transformation $\tau(k)$. For an example,
see Figure~\ref{fig:probes}(a,b).
\begin{lemma}
$\mu^* = \max_{0 \leq i < k < n} \mu^*(B_{ik})$.
\end{lemma}
\begin{proof}
For any $x < \mu^*$, let $a, b$ be any pair such that $a \prec b$ and
$\mu(C(a,b)) > x$. Now let $c$ be the longest common prefix of $a$ and $b$. Note that
$a$ and $b$ have the same number of digits, so $c$ must be a proper prefix of
both $a$ and $b$.
Let $\hat{a}$ be the digit of $a$ following the prefix $c$, and
let $\hat{b}$ be the digit of $b$ following the prefix $c$.
Thus $C(a)$ lies in subregion $\hat{a}$
of $C(c)$, and $C(b)$ lies in subregion $\hat{b}$ of $C(c)$. Since $\mu$ is
invariant under $\tau(c)$, it follows that
$\mu^*(B_{\hat{a}\hat{b}}) \geq \mu(C(a,b)) > x$.
Thus, for any $x < \mu^*$, there are $0 \leq i < k < n$
such that $\mu^*(B_{ik}) > x$. This implies there are
$0 \leq i < k < n$ such that $\mu^* \leq \mu^*(B_{ik})$, which proves the lemma.
\end{proof}

Let \emph{refinement} $\refinement_{ij}(P)$ of probe $P$, with $i, j \in \{0,...,n-1\}$, be the probe
with front transformation $\alpha(P) \circ \tau(i)$,
tail transformation $\omega(P) \circ \tau(j)$,
and midsection $\alpha(P) (C(\succ i)) \cup M(P) \cup \omega(P) (C(\prec j))$;
see Figure~\ref{fig:probes}(c,d).
Since $P = \bigcup \refinement_{ij}(P)$, we have $\mu^*(P) = \max \mu^*(\refinement_{ij}(P))$.

We say a probe $P$ is in canonical form if $\alpha(P)$ is the identity transformation.
We can construct a canonical form $\canon$ of any probe $P$ by setting $\alpha(\canon)$ to
the identity transformation, $M(\canon) := \alpha(P)^{-1}(M(P))$, and
$\omega(\canon) := \alpha(P)^{-1} \circ \omega(P)$; see Figure~\ref{fig:probes}(c,e).
Since $\mu$ is invariant under all
transformations involved, we have $\mu^*(P) = \mu^*(\canon)$.
Therefore it suffices to work only with probes in canonical form, where
the \emph{children} of a canonical probe $\canon$ are the canonical forms of its
refinements. So child $\canon_{ij}$ is the canonical probe with midsection
$M(\canon_{ij}) := \tau(i)^{-1} (C(\succ i) \cup M(\canon) \cup \omega(\canon) (C(\prec j)))$
and tail transformation $\omega(\canon_{ij}) := \tau(i)^{-1} \circ \omega(\canon) \circ \tau(j)$ (Figure~\ref{fig:probes}(e)).
Observe that $\tau(i)^{-1}$ always includes scaling with a factor greater than 1,
so for any canonical probe $\canon$ and any canonical child $\canon_{ij}$ we have
$|M(\canon_{ij})| > |M(\canon)|$, unless $i = n-1$, $M(\canon) = \emptyset$ and $j = 0$.

Note that while computing $\mu^*(\canon)$ may be difficult, it may be easy to get
a lower bound $\mu^-(\canon)$ and an upper bound $\mu^+(\canon)$ on $\mu^*(\canon)$.
For example, if $\mu(A)$ is defined as $|\mathrm{bbox}(A)| / |A|$, then
$|\mathrm{bbox}(M(\canon))| / |C \cup M(\canon) \cup \omega(\canon)(C)|$
would be a lower bound on $\mu^*(\canon)$, and
$|\mathrm{bbox}(C \cup M(\canon) \cup \omega(\canon)(C))| / |M(\canon)|$
would be an upper bound on $\mu^*(\canon)$ (provided $|M(\canon)| > 0$, otherwise we define $\mu^*(\canon) := \infty$).

\subsection{Searching probes}

\begin{figure}
\fbox{\vbox{\hsize=0.95\hsize\hrule height0pt\vskip-0.75\baselineskip\hrule height0pt\begin{codebox}
\Procname{$\proc{ComputeCurveQuality}$}
\li $Q \gets$ an empty first-in-first-out queue
\li $R \gets$ an empty dictionary
\li Insert the canonical forms of all base probes $B_{ik}$ in $Q$ and in $R$ \label{initQ}
\li $\id{lowerBound} \gets \max_{\canon \in Q} \mu^-(\canon)$
\li \While we do not like the gap between $\id{lowerBound}$ and $\max_{\canon \in Q} \mu^+(\canon)$
\li \Do  Extract a probe $\canon$ from the head of $Q$
\li      \For all canonical children $\canon_{ij}$ of $\canon$
\li      \Do  \If $\mu^+(\canon_{ij}) \geq \id{lowerBound}$ and $R$ does not contain $\canon_{ij}$ \label{checkR}
\li           \Then Add $\canon_{ij}$ to $Q$ and $R$
\li                 $\id{lowerBound} \gets \max(\id{lowerBound}, \mu^-(\canon_{ij}))$
              \End
         \End
    \End
\li Report that $\mu^*$ lies in the interval $[\id{lowerBound},\max_{\canon \in Q} \mu^+(\canon)]$.
\end{codebox}\vskip-\baselineskip}}
\caption{Algorithm to compute an approximation of a curve quality measure.}
\label{fig:algorithm}
\end{figure}

Our general algorithm to approximate $\mu^*$ is shown in Figure~\ref{fig:algorithm}.
The main idea of the algorithm is that it keeps replacing probes by their refinements
to get tighter lower and upper bounds on~$\mu^*$.
It is easy to see that the algorithm produces a correct lower bound.
We will prove that the algorithm also produces a correct upper bound
by proving that the following invariant holds after every iteration of the while loop:
for every probe $\canon$ that was ever added to the queue and has
$\mu^*(\canon) = \mu^*$, there is a descendant $\canon'$ of $\canon$ in $Q$ such that $\mu^*(\canon') = \mu^*$.

In fact the algorithm would be trivial to prove correct if we would
omit the check if $\canon_{ij}$ is contained in $R$ on line~\ref{checkR}. However, the
algorithm would be useless, because the queue would continue to contain \emph{degenerate probes}, that is,
probes $\canon$ with $M(\canon) = \emptyset$. This is because some degenerate probes are
inserted on line~\ref{initQ} (namely all base probes $B_{ik}$ with $k = i+1$), and whenever
a degenerate probe $\canon$ is extracted from the queue, its degenerate child
$\canon_{n-1,0}$ would be added to the queue. Since $\mu^+(\canon) = \infty$ for
degenerate probes, the algorithm would never find an actual upper bound on $\mu^*$.

Fortunately, it is easy to see that for most space-filling curves the algorithm can
only generate a small number of \emph{different} degenerate probes: typically
for any degenerate canonical probe the transformation $\omega$ has scale factor one, rotations
and reflections form a small closed set, and the translation is fixed because the
tail section has to connect to the front section. Moreover, non-degenerate probes
have only non-degenerate children.
Therefore, making sure that no probe is inserted
in $Q$ more than once, guarantees that $Q$ soon becomes and remains
free of degenerate probes, and the algorithm soon finds an upper bound on $\mu^*$.
We have to prove, however, that this upper bound is indeed correct.

\begin{theorem}
Algorithm $\proc{ComputeCurveQuality}$ returns correct lower and upper bounds on~$\mu^*$.
\end{theorem}
\begin{proof}
We need to prove that, despite the fact that the algorithm refuses to insert probes
in $Q$ that were inserted before, the invariant still holds after every iteration:
for every probe $\canon$ that was ever added to the queue and has $\mu^*(\canon) = \mu^*$,
there is a descendant $\canon'$ of $\canon$ in $Q$ such that $\mu^*(\canon') = \mu^*$.

For the sake of contradiction, let $\canon$ be the probe that is removed from $Q$ in
the first iteration that violates the invariant and does not restore it by inserting
children of $\canon$. Since $\mu^*(\canon) = \mu^*$, the probe $\canon$ must have a child $\canon_{ij}$
with $\mu^*(\canon_{ij}) = \mu^* \geq \id{lowerBound}$. If this child is not inserted
in $Q$ (restoring the invariant), it must be because $R$ contains $\canon_{ij}$ already,
which implies that at the beginning of the iteration $Q$ contained a descendant $\canon''$
of $\canon_{ij}$ with $\mu^*(\canon'') = \mu^*$. Since $\canon''$ is also a descendant of $\canon$, the
invariant that $Q$ holds a descendant $\canon'$ of $\canon$ with $\mu^*(\canon') = \mu^*$ can only
be violated by also removing (at least) $\canon''$. But we remove only one probe from
the queue, so this implies $\canon'' = \canon$, and we have that $\canon_{ij}$ is a descendant
of $\canon$ and vice versa. Now as observed above, from canonical parent to canonical
child the size of the midsection is either zero or strictly increasing. So if $\canon$ is
a descendant of $\canon_{ij}$, we must have $M(\canon) = M(\canon_{ij}) = \emptyset$, $i = n-1$,
and $j = 0$.

Consider a full line of ancestry of $\canon_{ij}$ down from itself, that is, a sequence
of canonical probes $\canon^0, ..., \canon^k$, where $\canon^0 = \canon_{ij}, \canon^{k-1} = \canon$, $\canon^k = \canon^0$,
and each probe $\canon^{m+1}$ is a child of $\canon^m$, for $0 \leq m < k$.
We have $\mu^*(\canon^m) = \mu^*$ for all $0 \leq m \leq k$.
We prove by induction on $m$ that at the end of the current iteration, we have for all $0 \leq m < k$ that $\canon^m$ is in
$R$ but not in $Q$. For $\canon^0 = \canon_{ij}$
this is true because the violation of the invariant is caused by not inserting $\canon_{ij}$ in $Q$,
which must be because it is in $R$ already.
Now, given that $\canon^0, ..., \canon^{m}$ are in $R$ but not in $Q$, consider $\canon^{m+1}$.
The probe $\canon^{m+1}$ is in $R$, because $\canon^{m}$ is in $R$ and not in $Q$; when $\canon^m$
was removed from $Q$, its child $\canon^{m+1}$, with $\mu^*(\canon^{m+1}) = \mu^* \geq \id{lowerBound}$,
must have been added to $R$ if it was not in $R$ already. But again $\canon^{m+1}$ is
not in $Q$, otherwise $Q$ would contain a descendant of $\canon$ and the invariant would
not be violated.

Therefore $\canon^0, ..., \canon^k$ are all in $R$ but not in $Q$, that is, they were all
once added to the queue and have been extracted since. It follows that any
non-degenerate children $\canon^*$ of $\canon^0, ..., \canon^k$ with $\mu^*(\canon^*) = \mu^*$
were once inserted in $R$ and $Q$, so $Q$ must
still contain descendants of any such non-degenerate children $\canon^*$.
The removal of $\canon$, a degenerate probe, did not change that, so if there was
ever such a non-degenerate child $\canon^*$ with $\mu^*(\canon^*) = \mu^*$, then a descendant $\canon'$ of $\canon^*$,
and thus, of $\canon$, with $\mu^*(\canon') = \mu^*$, would still exist in $Q$ and the
invariant would not be violated. So we must conclude that for $0 \leq m < k$,
the degenerate probe $\canon^{m+1}$ is the \emph{only} child $\canon^*$ of $\canon^m$ with
$\mu^*(\canon^*) = \mu^*$. It follows that $\canon$ in particular does not have any
non-degenerate descendant probes $\canon^*$ with $\mu^*(\canon^*) = \mu^*$.

Now, for any value $x < \mu^*$, let $a, b$ be any pair such that $\mu(\canon(a,b)) > x$.
Let $\hat{a}$ be $a$ followed by a zero, and let $\hat{b}$ be $b$ followed by a zero. Now
we have $\mu(\canon(\hat{a},\hat{b})) = \mu(\canon(a,b)) > x$, and $C(\succ \hat{a}) \neq \emptyset$.
The probe $\canon^*$ with midsection
$\tau(a)^{-1} (C(\succ \hat{a}) \cup M(\canon) \cup \omega(\canon) (C(\prec \hat{b})))$,
and tail transformation $\tau(a)^{-1} \circ \omega(\canon) \circ \tau(\hat{b})$,
is a non-degenerate descendant probe of $\canon$ with $\mu(\canon^*) > x$. Since we can
find such a non-degenerate descendant probe of $\canon$ for every $x < \mu^*$, this
contradicts the conclusion of the previous paragraph that
$\canon$ does not have any non-degenerate descendant probes $\canon^*$ with $\mu^*(\canon^*) = \mu^*$.

Hence there cannot be such a probe $\canon$ whose removal from $Q$ leads to the invariant
being violated at the end of an iteration. Therefore the invariant is maintained,
and the algorithm is correct.
\end{proof}

\subsection{Implementation}
\label{sec:implementation}

To implement algorithm $\proc{ComputeCurveQuality}$ for a particular measure, one needs to come
up with a representation of midsections of canonical probes $\canon$ that enables a quick,
correct, and reasonably tight approximation of $\mu^*(\canon)$ from above and below.
If the approximations are tight, the algorithm may zoom in on the worst-case sections of the
curve quickly, keeping a small probe queue~$Q$ and refining only where necessary. If the
approximations are not tight enough, the algorithm may degenerate into a search of a
complete grid, taking long to converge or not converging at all. For an efficient operation
of the algorithm the representation of the midsections should also be easy to maintain under
the transformations in the rules that describe the scanning order.

For some curves and measures a good implementation is easier to accomplish than for others.
To compute \WXD, \WED, and \WRD we only need to know the size (area) of the midsection;
for \WBA and \WBP we also need to know the minimum rectangular bounding box of the midsection;
for \WOA and \WOP we need to know the minimum octagonal bounding box. For curves that only
use axis-parallel reflections and rotations with angles of $\frac12\pi$, $\pi$ and $\frac32\pi$,
these midsection properties can be maintained easily. Fortunately, most curves presented in
Section~\ref{thecurves} fulfill these requirements.
However, the Sierpi\'nski-Knopp order and the Gosper flowsnake use rotations that are not
multiples of $\pi/2$, and as a result the \WXD, \WRD, \WBA, \WBP, \WOA and \WOP measures are
not invariant under rotation as required by the algorithm.

\begin{figure}[tb]
  \centering
  \includegraphics[width=0.65\hsize]{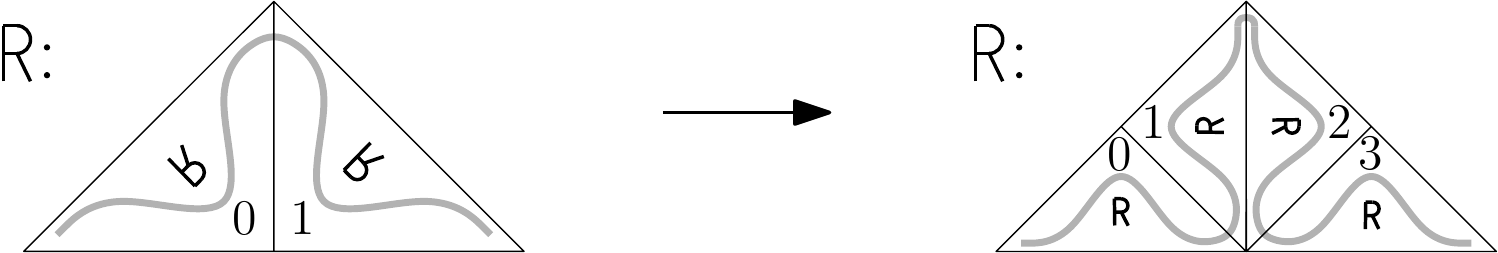}
  \caption{Expanding the ordering rule of the Sierpi\'nski-Knopp order by one level results in
           rotations by multiples of $\pi/2$.}
  \label{fig:sierpinski-expansion}
\end{figure}
For the Sierpi\'nski-Knopp order this is easily solved by expanding the definition by one
level of recursion, see Figure~\ref{fig:sierpinski-expansion}.

For the Gosper flowsnake we observe that \WED is invariant under rotation and can still be
computed by our algorithm. Because the rotations of the curve are not a rational fraction of $\pi$,
any pair of points that defines the worst-case $L_2$-locality will appear scaled down and turned
arbitrarily close to horizontal somewhere in the curve, so that we have $\WXD \geq \WED$.
Since the $L_2$-distance between any pair of points is at least their $L_\infty$-distance, it
follows that $\WXD = \WED$.
Likewise, any pair of points that defines the worst-case $L_2$-locality will appear scaled down
and turned arbitrarily close to making a 45 degree angle with the horizontal, so that we have
$\WRD \geq 2 \WED$ and $\WBP \geq \WBA \geq \WED/2$. Since the squared $L_1$-distance between any pair of points
is at most twice their squared $L_2$-distance, it follows that $\WRD = 2\WED$.
For the computation of $\WBA$, $\WBP$, $\WOA$ and $\WOP$ similar arguments could be used, provided
the boundaries of unit regions and midsections are represented in such a way that one can determine
good lower and upper bounds on the maximum, over all possible rotations, of the size of the minimum
bounding box. This seems hard to accomplish due to the fractalic nature of the region boundaries.

\section{Computational results}
\label{results}

\paragraph{Computing pairwise worst-case measures}
We implemented the approximation algorithm described above, specifically to compute
the worst-case locality measures $\WXD$, $\WED$, and $\WRD$, the worst-case bounding box
quality measures $\WBA$ and $\WBP$, and the corresponding measures for bounding octagons $\WOA$ and $\WOP$.
We ran our algorithm on the curves mentioned in Section~\ref{thecurves}, except that
we did not try to compute $\WBA$, $\WBP$, $\WOA$ and $\WOP$ for the Gosper flowsnake,
because of the reasons explained in Section~\ref{sec:implementation}.

The running times of the computations varied with the curves. When the algorithm succeeded in
having a probe queue of constant size, we would have an approximation with precision 0.0001 within a
fraction of a second (higher precision being prevented by the number of bits used to
represent numbers); when the probe queue kept growing, the computation could take
a few minutes and the precision would be limited by memory requirements.
\WBA, \WBP, \WOA and \WOP were always computed fast and precise; \WRD was fast and
precise for all orders except Sierpi\'nski-Knopp order; \WED and \WXD were generally computed only
to a precision of 0.0005.

We tested the 278 different orders that fit Wunderlich's scheme and are defined on a
$2 \times 2$-grid (that is, Hilbert order), a $3 \times 3$-grid (R-order and the 272
Serpentine orders), or a $4 \times 4$-grid (4~different orders, not counting Hilbert order).
From these orders five curves ${\cal C}$ turned out to be ``dominant'' in the
sense that there was no curve that was better than ${\cal C}$ on at least one measure and at least as good
as ${\cal C}$ on the other measures. These five curves are the Hilbert order
(best on \WRD, \WBP and \WOP),
Serpentine 000\,000\,000 (that is, GP-order, best on \WBA and \WOA),
Serpentine 011\,010\,110 and 101\,010\,101 (with equal scores, not best on anything but
not dominated either), and Serpentine 110\,110\,110 (Meurthe, best on \WED and \WXD).

\paragraph{Computing average total measures}
We examined the five dominant curves further, together with coil order, balanced GP-order,
$\beta\Omega$-order, \ARRWW-order, Z-order and Sierpi\'nski-Knopp order. For these curves we estimated
the average total bounding box area and the square average relative total bounding box perimeter
(also for bounding octagons)
by random sampling as follows: we generated 100 sets of
numbers chosen uniformly between 0 and 1 that subdivide the curve, where the logarithm
of the size of each set was chosen uniformly between $\log 500$ and $\log 18\,000$.
Thus we cover an integral number of periods of fluctuation for each scanning order.
We estimated the average total area (or relative perimeter) by taking the average over
these 100 sample subdivisions.
Because in some applications it is useful to keep the
$L_1$-diameter of curve sections small and our software was easy to adapt to it, we
used the same method to also estimate the
square average relative total curve section diameter in the $L_\infty$ metric (\AXD)
and in the $L_1$ metric (\ARD).

\paragraph{Results}
The results of our computations are shown in Table~\ref{tbl:bounds}.
Note that for some
scanning orders the exact worst-case locality measures were already known:
tight lower and upper bounds for the Hilbert order were proven one by one in several
papers: lower bounds on \WXD, \WED, and \WRD by Alber and Niedermeier~\cite{alber},
Gotsman and Lindenbaum~\cite{gotsman}, and Niedermeier and Sanders~\cite{niedermeier-manhattan},
respectively; upper bounds by Bauman~\cite{bauman}, Bauman~\cite{bauman}, and Chochia et al.~\cite{chochia},
respectively. Tight bounds for GP-order, coil order (Luxburg 1), and Luxburg 2 were proven
by Luxburg~\cite{luxburg}, and tight bounds for Sierpi\'nski-Knopp order were proven by
Niedermeier et al.~\cite{niedermeier}, confirming earlier observations on \WED~\cite{mandelbrot}
(with respect to these measures, the H-order described by Niedermeier et al.\ is equivalent to
Sierpi\'nski-Knopp order and to Ces\`aro's variant of the Von Koch curve as mentioned by
Mandelbrot~\cite{mandelbrot}).
Our computations confirmed all of these results.
The other bounds have been computed by us. The bounds on the W-measures are the average of lower
and upper bounds which have a gap of at most 0.0005;
all printed numbers are less than 0.001 off from the real values.
Only the bounds for the Gosper flowsnake are less precise (this order involves
rotations by angles of $\arctan \frac15\sqrt 3$, which
makes it more challenging to get bounds with high precision).

The bounds on the A-measures result from our experiments with random subdivisions of curves.
We omit the results for \ARD, because
for all curves we found that $2 \AXD \leq \ARD$. This means that the best total
$L_1$-diameter is obtained by rotating the curve by 45 degrees, so that the
square total $L_1$-diameter becomes twice the original square total $L_\infty$-diameter.
The Sierpi\'nski-Knopp order was the only one not affected by the rotation.

\begin{table}[tbp]
  \centering
  \small
  \begin{tabularx}{\hsize}{|@{\,}l@{\,}|@{\,}r@{}X@{}r@{}X@{}r@{}X@{}||@{\,}r@{}X@{}@{ }r@{}X@{}|@{\,}r@{}X@{}@{ }r@{}X@{}||@{\,}r@{}X@{}@{ }r@{}X@{}|@{\,}r@{}X@{}||@{\,}r@{}X@{}|}
  \hline
  Order & \multicolumn{2}{@{}l@{}}{\WXD}
        & \multicolumn{2}{@{}l@{}}{\WED}
        & \multicolumn{2}{@{}l@{}||}{\WRD}
        & \multicolumn{2}{@{}l@{}}{\WBA}
        & \multicolumn{2}{@{}l@{}|}{\ABA}
        & \multicolumn{2}{@{}l@{}}{\WBP}
        & \multicolumn{2}{@{}l@{}||}{\ABP}
        & \multicolumn{2}{@{}l@{}}{\WOA}
        & \multicolumn{2}{@{}l@{}|}{\AOA}
        & \multicolumn{2}{@{}l@{}||}{\WOP}
        & \multicolumn{2}{@{}l@{}|}{\AXD} \\
  \hline
  Sierpi\'nski-Knopp order                    & 4&      & 4&      & 8&      & 3&.000  & 1&.78    & 3&.000  & 1&.42   & 1&.789  & 1&.25   & 1&.629  & 1&.77$'$\\
  \textbf{Balanced GP}               & 4&.619  & 4&.619  & 8&.619  & 2&.000  & 1&.44    & 2&.155  & 1&.19   & 1&.769  & 1&.31   & 1&.807  & 1&.72$'$\\
  \hline
  GP (Serp.\,000\,000\,000)          & 8&      & 8&      &10&\twoth& 2&.000  & 1&.44    & 2&.722  & 1&.28   & 1&.835  & 1&.32   & 2&.395  & 2&.13$'$\\
  \textbf{Serpentine\,011\,010\,110} & 5&.625  & 6&.250  &10&.000  & 2&.500  & 1&.44$''$& 2&.500  & 1&.20   & 2&.222  & 1&.32$'$& 2&.036  & 1&.71$'$\\
  Luxburg 2 (101\,010\,101)          & 5&\fivee& 6&\quart&10&      & 2&.500  & 1&.49$'$ & 2&.500  & 1&.24   & 2&.222  & 1&.35$'$& 2&.036  & 1&.81$'$\\
  \textbf{Meurthe (110\,110\,110)}   & 5&.333  & 5&.667  &10&.667  & 2&.500  & 1&.41$''$& 2&.667  & 1&.17   & 2&.000  & 1&.30$'$& 2&.018  & 1&.64$'$\\
  Coil (Serp.\,111\,111\,111)        & 6&\twoth& 6&\twoth&10&\twoth& 2&.500  & 1&.41$'$ & 2&.667  & 1&.17   & 2&.222  & 1&.29$'$& 2&.424  & 1&.63$'$\\
  \hline
  Hilbert                            & 6&      & 6&      & 9&      & 2&.400  & 1&.44    & 2&.400  & 1&.19   & 1&.929  & 1&.30   & 1&.955  & 1&.67$'$\\
  $\beta\Omega$                      & 5&.000  & 5&.000  & 9&.000  & 2&.222  & 1&.42    & 2&.250  & 1&.17   & 1&.800  & 1&.29   & 1&.933  & 1&.64$'$\\
  \ARRWW                             & 5&.400  & 6&.046  &12&.000  & 3&.055  & 1&.49$'$ & 3&.125  & 1&.22   & 2&.344  & 1&.33   & 2&.255  & 1&.70$'$\\
  Z-order                            &&$\infty$&&$\infty$&&$\infty$&&$\infty$& 2&.92    &&$\infty$& 2&.40$'$&&$\infty$& 2&.46   &&$\infty$& 3&.80$''$\\
  \hline
  Gosper flowsnake                   & 6&.35   & 6&.35   &12&.70 &\multicolumn{2}{@{}l@{}}{$\geq$3.18}&  &     &\multicolumn{2}{@{}l@{}}{$\geq$3.18}&  &      &  &      &  &      &  &      &  &   \\
  \hline
  \end{tabularx}
  \caption{Bounds for different measures and curves. New curves printed in bold. For the A-measures
           the standard deviation is indicated behind the number: no symbol when less than 0.5\%;
           one mark when between 0.5\% and 1.0\%, two marks when between 1.0\% and 2.0\%.}
  \label{tbl:bounds}
\end{table}

Regarding worst-case locality in the $L_\infty$- and $L_2$-metrics, we see that the best order in Wunderlich's scheme
had not yet been found: Meurthe (Serpentine 110\,110\,110) turns out to have even better locality in these measures
than Luxburg's second variant (Serpentine 101\,010\,101). Even better locality is achieved by Wierum's $\beta\Omega$-curve
(matching or improving on Hilbert's curve in all measures) and still better by our new Peano variant: balanced GP\@.
The latter approaches the locality of the Sierpi\'nski-Knopp order, which is still conjectured to be optimal.

However, it appears that the optimal locality of the Sierpi\'nski-Knopp order comes at a price: it results in high worst-case bounding
box measures, and in our experiments on random subdivisions the resulting bounding boxes are about 25\% worse than with most other orders.
Only Z-order, which tends to result in bounding boxes twice as big as with the other orders, performs worse.
The best performer on the worst-case bounding box measures is our balanced GP-order, which also performs
well in the experiments on random subdivisions (similar to Hilbert), but coil order, Meurthe order and $\beta\Omega$-order perform even better
in the experiments.

We also computed worst-case and average total bounding box measures for rectangular bounding boxes with edges at 45 degrees'
angles with the coordinate axes. This was (very) harmful with all scanning orders except Sierpi\'nski-Knopp order, where it had no effect
(this was to be expected, because the definition of Sierpi\'nski-Knopp order involves rotations by all multiples of 45 degrees).
The Sierpi\'nski-Knopp order benefits more than any other from using octagonal bounding boxes instead of
rectangular bounding boxes, as we can see in the right columns: here Sierpi\'nski-Knopp order is the best performer. However, all things
considered the advantages of bounding octagons may not be worth the effort. Such octagonal bounding boxes need
twice the description size of rectangular bounding boxes, but the savings are small: in total bounding octagons
constructed by Sierpi\'nski-Knopp order are only 13\% smaller than bounding rectangles constructed by Hilbert or GP-order.

\section{Conclusions}

\paragraph{Pairwise worst-case measures}
Known locality measures of space-filling curves do not predict
well how effective they are when used to group points into bounding boxes. Therefore
we proposed new measures of bounding-box quality of space-filling curves. We presented
new scanning orders that perform well on these measures, most notably the
\emph{balanced GP-order}, which has worst-case bounding box area ratio (\WBA) 2.000, and
worst-case bounding box square perimeter ratio (\WBP) 2.155. On worst-case locality measures this
curve also scores very well, much better than Peano's original curve, and beaten only by
Sierpi\'nski-Knopp order.

We conjecture that a worst-case bounding box area ratio (\WBA) of 2 is in fact optimal and
cannot be improved by any (recursively defined) space-filling curve. More provocatively we
conjecture that the optimal worst-case bounding box square perimeter ratio (\WBP) is also~2
(note that we have not actually found a curve with \WBP less than 2.155).
We add these conjectures to those by Niedermeier et al., who conjectured that the optimal
\WXD, \WED, and \WRD locality values are 4, 4, and 8, respectively (Niedermeier et al.\ posed this conjecture
for curves filling a square, but we would like to drop this restriction).

For proving the lower bounds in these conjectures we have come a long way. Niedermeier et al.\ proved
tight lower bounds on the worst-case locality for a certain class of space-filling curves.
Unfortunately, strictly speaking almost none of the space-filling curves in our study belongs to
that class. For $L_2$-locality and for worst-case bounding-box area and squared perimeter,
we managed to prove the conjectured lower bounds for another class of space-filling curves,
partly overlapping with the class covered by Niedermeier et al., and now including almost all
space-filling curves mentioned in this paper.

\begin{figure}[tb]
  \centering
  \includegraphics[width=\hsize]{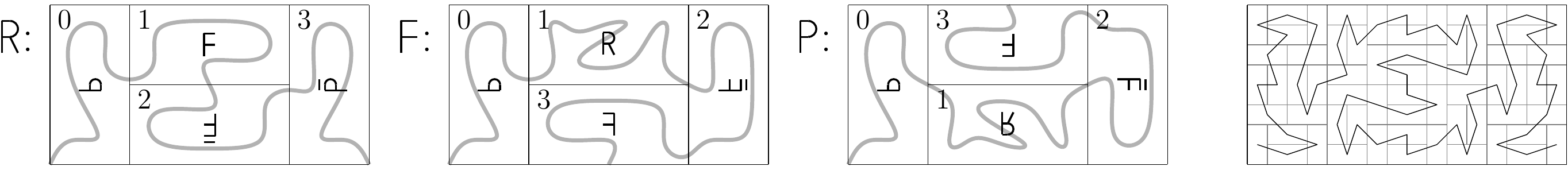}
  \caption{Example of a scanning order based on rectangles, but not in a standard grid pattern.
           The locality and bounding-box quality of this particular scanning order are not very
           interesting: $\WXD = 6.000$, $\WED = 6.667$, $\WRD = 12.656$, $\WBA = 3.125$, $\WBP = 3.164$.}
  \label{fig:bricks}
\end{figure}
Still we have not been able to prove these lower bounds for \emph{all} space-filling curves.
Our proof is restricted to scanning orders based on axis-aligned rectangles in a regular
grid pattern and to scanning orders based on triangles.
In this paper we mentioned the Gosper flowsnake curve, which is based on a
subdivision of a unit region into subregions that all have fractalic boundaries, and is
therefore not included in the class of scanning orders to which our lower bounds apply.
Of course one may argue that from a practical point of view it is questionable if one
would like to use such a curve anyway, but we can also come up with scanning orders based
on tiling the plane with rectangles---but not in a standard grid pattern---or L-shapes, for example
(see Figure~\ref{fig:bricks}).
Our bounds do not apply to such curves, and the bounds by Niedermeier et al.\ do not
necessarily apply either.

Regarding lower bounds on $L_1$-locality we did not make any progress on Niedermeier et al.


\paragraph{Performance in practice(?)}
Our experiments on random points may give an impression of how effective the different
curves would be in the application considered in this paper: a data structure for points in the plane,
based on sorting the points into blocks of points that are consecutive along the curve.
We see that it would be clearly suboptimal to use the order with the best
$\WD_\infty, \WD_2$ and $\WD_1$ locality (Sierpi\'nski-Knopp order) for this application.
It seems to be better indeed to choose a curve based on \WBA and \WBP (balanced GP-order).

Still the \WBA and \WBP measures do not predict performance on random points perfectly
either: there are several curves with only moderate \WBA and \WBP values that seem to be
as effective as the balanced GP-order (for example Hilbert order) or even slightly better
(for example coil order or $\beta\Omega$-order) on random point data. It should also be
noted that total bounding-box area and perimeter may not be the only factors that determine
the performance of a curve in a data structure setting. Asano et al.~\cite{asano} argued that
in certain settings it would be good if any axis-parallel square query window can be covered
by a small number of contiguous sections of the curve that together cover an area of at
most a constant times that of the query window. It would be interesting to have an algorithm
that can analyse a given curve automatically to compute the constants involved, in the
worst case and on average.

\paragraph{Higher dimensions}
For what the \WBA and \WBP measures are worth, the conjectured near-optimality of the
balanced GP-order suggests that there is little room for hope to find significantly more
effective scanning orders in two dimensions. A first topic for further research is to
determine the gap between our lower bound constructions and the performance of known
space-filling curves when we consider generalisations to three dimensions. Chochia and Cole~\cite{chochia}
and Niedermeier et al.~\cite{niedermeier} have some results on
locality, but the gap is large and the field is still wide open,
especially with respect to bounding box quality.

Still higher
dimensions can be interesting; four-dimensional space-filling curves can be particularly
interesting to order rectangles in the plane (which are specified by four coordinates
each)~\cite{arge,haverkort,kamel1993}. A first challenge in that context is to define
appropriate quality measures that say something sensible about the quality of bounding
boxes in two dimensions that are formed by grouping points in four dimensions.


\small
\bibliographystyle{abbrv}

\begin{thebibliography}{10}

\bibitem{alber}
J.~Alber and R.~Niedermeier.
\newblock On multidimensional curves with {H}ilbert property.
\newblock {\em Theory of Computing Systems}, 33(4):295--312, 2000.

\bibitem{akiyama}
J.~Akiyama, H.~Fukuda, H.~Ito, and G.~Nakamura.
\newblock Infinite series of generalized Gosper space filling curves.
\newblock In {\em China-Japan Conf. on Discrete Geometry, Combinatorics and Graph Theory 2005}, LNCS 4381, pp 1--9, 2007.

\bibitem{arge}
L.~Arge, M.~{de Berg}, H.~J. Haverkort, and K.~Yi.
\newblock The {P}riority {R}-tree: a practically efficient and worst-case
  optimal {R}-tree.
\newblock {\em ACM Transactions on Algorithms}, 4(1):9, 2008.

\bibitem{asano}
T. Asano, D. Ranjan, T. Roos, E. Welzl, P. Widmayer:
Space-Filling Curves and Their Use in the Design of Geometric Data Structures.
{\em Theoretical Computer Science} 181(1):3--15, 1997.

\bibitem{bauman}
K.~E. Bauman.
\newblock The dilation factor of the {P}eano-{H}ilbert curve.
\newblock {\em Math. Notes}, 80(5):609--620, 2006.

\bibitem{chochia}
G.~Chochia, M.~Cole, and T.~Heywood.
\newblock Implementing the hierarchical {PRAM} on the {2D} mesh: Analyses and
  experiments.
\newblock In {\em Symp. on Parallel and Distributed Processing}, 1995, pp 587--595.

\bibitem{faloutsos}
C. Faloutsos, S. Roseman:
Fractals for Secondary Key Retrieval.
In {\em Symp. on Principles of Database Systems}, 1989, pp 247--252.

\bibitem{fishburn}
P. Fishburn, P. Tetali, P. Winkler:
Optimal Linear Arrangement of a Rectangular Grid.
{\em Discrete Math.} 213:123--139, 2000.

\bibitem{gardner}
M.~Gardner.
\newblock Mathematical Games---In which ``monster'' curves force redefinition of the word ``curve''.
\newblock {\em Scientific American}, 235(6):124--133, 1976.

\bibitem{gotsman}
C.~Gotsman and M.~Lindenbaum.
\newblock On the metric properties of discrete space-filling curves.
\newblock {\em IEEE Trans. Image Processing}, 5(5):794--797, 1996.

\bibitem{haverkort}
H.~Haverkort and F.~{van Walderveen}.
\newblock {\em Four-dimensional Hilbert curves for R-trees}.
\newblock Manuscript submitted to Int. Symp. on Algorithms and Computation (ISAAC), 2008.

\bibitem{hilbert}
D.~Hilbert.
\newblock {\"U}ber die stetige {A}bbildung einer {L}inie auf ein
  {F}l{\"a}chenst{\"u}ck.
\newblock {\em Math. Ann.}, 38(3):459--460, 1891.

\bibitem{hungershoefer}
J. Hungersh\"ofer, J.-M. Wierum:
On the Quality of Partitions Based on Space-Filling Curves.
In {\em Int. Conf. on Computational Science} 2002, LNCS 2331:36--45.

\bibitem{jagadish}
H. V. Jagadish:
Linear clustering of objects with multiple attributes.
In {\em ACM SIGMOD conf. on Management of Data} 1990, pp 332--342.

\bibitem{kamel1993}
I.~Kamel and C.~Faloutsos.
\newblock On packing {R}-trees.
\newblock In {\em Conf. on Information and Knowledge Management} 1993, pp 490--499.

\bibitem{lebesgue}
H.~L. Lebesgue.
\newblock {\em Le\c cons sur l'int{\'e}gration et la recherche des fonctions
  primitives}, pp 44--45.
\newblock Gauthier-Villars, 1904.

\bibitem{luxburg}
U.~{von Luxburg}.
\newblock {\em Lokalit{\"a}tsma{\ss}e von Peanokurven}.
\newblock Student project report, Universit{\"a}t T{\"u}bingen, Wilhelm-Schickard-Institut f{\"u}r Informatik, 1998.

\bibitem{mandelbrot}
B.~B.~Mandelbrot.
\newblock {\em The fractal geometry of nature}, p65.
\newblock W.~H.~Freeman and company, 1982.

\bibitem{manolopoulos}
Y.~Manolopoulos, A.~Nanopoulos, A.~N. Papadopoulos, Y.~Theodoridis.
\newblock {\em R-trees: Theory and Applications.}
Springer, 2005.

\bibitem{mitchison}
G. Mitchison, R. Durbin:
Optimal Numberings of an $N \times N$ Array.
{\em Siam J. Alg. Disc. Meth.} 7(4):571--582, 1986.

\bibitem{moon}
B. K. Moon, H. V. Jagadish, C. Faloutsos, J. H. Saltz:
Analysis of the clustering properties of the Hilbert space-filling curve.
{\em IEEE Trans. on Knowledge and Data Engineering}, 13(1):124--141, 2001.

\bibitem{moore}
E. H. Moore:
On certain crinkly curves,
{\em Trans. Amer. Math. Soc.}, 1 (1900), 72--90.

\bibitem{morton}
G.~M. Morton.
\newblock {\em A computer oriented geodetic data base and a new technique in file
  sequencing.}
\newblock Technical report, IBM, Ottawa, 1966.

\bibitem{niedermeier}
R.~Niedermeier, K.~Reinhardt, and P.~Sanders.
\newblock Towards optimal locality in mesh-indexings.
\newblock {\em Discrete Applied Mathematics}, 117:211--237, 2002.

\bibitem{niedermeier-manhattan}
R.~Niedermeier and P.~Sanders.
\newblock {\em On the {Manhattan}-distance between points on space-filling
  mesh-indexings.}
\newblock Technical Report IB 18/96, Karlsruhe University, Dept. of Computer
  Science, 1996.

\bibitem{peano}
G.~Peano.
\newblock Sur une courbe, qui remplit toute une aire plane.
\newblock {\em Math. Ann.}, 36(1):157--160, 1890.

\bibitem{sagan}
H.~Sagan:
{\em Space-Filling Curves.}
Universitext series, Springer, 1994.

\bibitem{wierum}
J.-M. Wierum.
\newblock {\em Definition of a new circular space-filling curve:
  {$\beta\Omega$}-indexing.}
\newblock Technical Report TR-001-02, Paderborn Center for Parallel Computing
  (PC$^2$), 2002.

\bibitem{wierum-logarithmic-path-length}
J.-M. Wierum:
Logarithmic Path-Length in Space-Filling Curves.
{\em Canadian Conference on Computational Geometry} 2002, pp 22--26

\bibitem{wunderlich}
W.~Wunderlich.
\newblock {\"U}ber {P}eano-{K}urven.
\newblock {\em Elemente der Mathematik}, 28(1):1--10, 1973.

\end{thebibliography}

\end{document}